\documentclass[11pt,english,longbibliography]{article}
\usepackage{url}

\usepackage{draft} 
\usepackage{putex}
\usepackage{comment}

\usepackage{hyperref}

\renewcommand{\vec}[1]{\mathbf{#1}}

\usepackage{amsthm}

\makeatletter
\newcommand*{\rom}[1]{\expandafter\@slowromancap\romannumeral #1@}
\makeatother

\usepackage[latin9]{inputenc}
\usepackage{multirow}
\usepackage{verbatim}
\usepackage{prettyref}
\usepackage[numbers,sort&compress]{natbib}
\usepackage{amsmath}
\usepackage{amssymb}
\usepackage{tikz-cd}
\tikzset{commutative diagrams/row sep/huge=4cm}
\tikzset{commutative diagrams/column sep/huge=4cm}
\tikzcdset{scale cd/.style={every label/.append style={scale=#1},
    cells={nodes={scale=#1}}}}
\usepackage{lmodern}
\usepackage{tcolorbox}
\usepackage{cases}
\usepackage{bbold}
\usepackage{bm}
\usepackage{ytableau}
\usepackage{youngtab}
\usepackage{mathtools}
\usepackage{graphicx}
\usepackage[nottoc]{tocbibind}
\usepackage{mathrsfs}
\usepackage{caption}
\usepackage{subcaption}
\usepackage{cancel}
\usepackage{float}
\usepackage{tikz}

\usetikzlibrary{arrows.meta}
\usetikzlibrary{bending}
\usepackage{tikz}

\tikzset{
    Witten diagram/.style={
        execute at begin picture={
            \draw[blue, line width=1.5pt] circle[radius=\pgfkeysvalueof{/tikz/Witten/radius}];
            \path node (X){\phantom{X}};
        },
        baseline={(X.base)}
    },
    vertex/.style={circle,fill,inner sep=1.5pt,node contents={}},
    Witten/.cd,
    radius/.initial=3cm
}
\usetikzlibrary{patterns}
\usepackage{lipsum}
\usepackage{framed}
\usepackage{adjustbox}
\usepackage{mathtools}
\usepackage{braket}

 \usepackage{slashed}
\usepackage{esvect}
\usepackage{dsfont}

\usepackage{color}
\definecolor{darkgreen}{rgb}{0,0.5,0}
\definecolor{darkblue}{rgb}{0,0,0.6}
\definecolor{purple}{rgb}{0.4,.2,0.7}

\numberwithin{equation}{section}
\numberwithin{figure}{section}
\numberwithin{table}{section}

\DeclareFontShape{OT1}{cmr}{mx}{n}{<->cmr10}{}

\begin{document}

\title{\centering Line Defects with a Cusp in Fermionic CFTs}

\authors{Simone Giombi \worksat{\PUJ}, Anurag Pendse \worksat{\PUJ}}

\institution{PUJ}{Joseph Henry Laboratories, Princeton University, Princeton, NJ 08544, USA}

\abstract{We study line defects with a cusp in fermionic CFTs arising as fixed points of scalar-fermion theories with Yukawa interactions. These include the Gross-Neveu-Yukawa model and some of its generalizations with additional scalar fields, which can be thought of as UV completions of fermionic theories with quartic interactions. We compute the cusp anomalous dimension in these models to one-loop order in the epsilon expansion near four dimensions, and also to leading order in the large $N$ expansion in $2<d<4$. We discuss several observables that can be extracted from the cusp anomalous dimension, such as the dimensions of the defect changing and defect creation operators, the Casimir energy appearing in the fusion of defects, and the normalization coefficients of the two-point functions of displacement and tilt operators. We provide some estimates of the values of these observables in $d=3$ using the one-loop epsilon expansion and Pad\'e approximants.}

\date{}

\maketitle

\tableofcontents

\section{Introduction and Summary}

In addition to local operators, quantum field theories often admit a variety of interesting extended objects, or defects. They have important applications in many contexts, such as gauge theory dynamics, non-perturbative dualities, generalized symmetries, holography, and condensed matter systems with boundaries, interfaces or impurities. In a conformal field theory (CFT), one may have defects that preserve a subgroup of conformal symmetries, yielding a defect conformal field theory (dCFT).  In this paper, we consider conformal line defects in CFTs with interactive fermions, including the Gross-Neveu universality class and some of its generalizations. We will mainly focus on the $d=4-\epsilon$ expansion, where these CFTs arise as fixed points of scalar-fermion theories with Yukawa interactions. As shown in \cite{giombi2025linedefectsfermioniccfts, Pannell_2023}, these models admit conformal line operators which are analogous to the so-called pinning field (or localized magnetic field) defect in the scalar $O(N)$ model \cite{Cuomo_2022}. Our goal here is to compute the cusp anomalous dimension associated to line defects with a cusp in these fermionic CFTs. While the cusp anomalous dimension has long been extensively studied in gauge theory \cite{Polyakov:1980ca, Korchemsky:1992xv}, its role in general critical systems has been less explored. A discussion of the theory of line defects with a cusp in CFT was recently given in \cite{cuomo2024impuritiescuspgeneraltheory}, where the explicit example of the pinning field defect in the $O(N)$ model was studied in some detail, using a combination of fuzzy sphere and epsilon expansion methods. 

One of the classic theories of interacting fermions is the Gross-Neveu (GN) model described by the action \cite{PhysRevD.10.3235} (we assume Euclidean signature throughout the paper)
\begin{equation}
    S=-\int d^dx\Big[\Bar{\Psi}\slashed{\partial}\Psi +\frac{g}{2}(\Bar{\Psi}\Psi)^2\Big]
\end{equation}
where 
\begin{equation}
    \Psi=\begin{pmatrix}
        \psi_1\, & \dots & \,\psi_{N_f}
    \end{pmatrix}
\end{equation}
denotes a collection of $N_f$ Dirac fermions. Below we will also use the notation $N=N_f c_d$, with $c_d$ being the number of components of a Dirac fermion in $d$ dimensions. This model is asymptotically free in $d=2$ for $N>2$, and hence in $d=2+\epsilon$ it has an interacting UV fixed point with $g\sim\epsilon$. This UV fixed point can also be studied using the large $N$ expansion in general $2<d<4$ (see section \ref{sec:largeN} below and \cite{Moshe_2003} for a review).  The GN model is believed to have a ``UV completion" given by a theory with one additional dynamical scalar field and Yukawa interactions \cite{HASENFRATZ199179, ZINNJUSTIN1991105}, known as Gross-Neveu-Yukawa theory
\begin{equation}
\label{GNY}
    S_{\rm GNY}=\int d^dx \Big[\frac{1}{2}(\partial\vec{\phi})^2-\Bar{\Psi}\slashed{\partial}\Psi + g_1 \phi\Bar{\Psi}\Psi +\frac{g_2}{24}\phi^4]\,.
\end{equation} 
This theory has perturbative IR fixed points in $d=4-\epsilon$ which are expected to be in the same universality class as the UV fixed points of the GN model. The additional scalar field in the GNY model is essentially related to the auxiliary field arising in the Hubbard-Stratonovich transformation of the quartic interaction in the GN model, which can be used to develop its large $N$ expansion (see section \ref{sec:largeN}). 

Another theory of a similar flavor is the Nambu Jona-Lasinio (NJL) model \cite{PhysRev.122.345} given by the action
\begin{equation}
S=-\int d^dx\Big[\Bar{\Psi}\slashed{\partial}\Psi+\frac{g}{2}\big[(\Bar{\Psi}\Psi)^2-(\Bar{\Psi}\gamma^5\Psi)^2\big]\Big]
\end{equation}
In addition to the $U(N_f)$ symmetry rotating the fermions, this model has a chiral $U(1)$ symmetry $\psi\rightarrow e^{i\gamma^5\alpha}\psi$ (the GN model, on the other hand, only has a discrete chiral symmetry). In $2<d<4$, it has UV fixed points that are again accessible by the large $N$ expansion, or by the epsilon expansion in $d=2+\epsilon$. Similarly to the GN model, the NJL model is expected to have a UV completion in terms of a theory with two scalar fields \cite{ZINNJUSTIN1991105}
\begin{equation}
S_{NJLY}= \int d^dx \Big[\frac{1}{2}(\partial\vec{\phi}_1)^2+\frac{1}{2}(\partial\vec{\phi}_2)^2-\Bar{\Psi}\slashed{\partial}\Psi +g_1 \Bar{\Psi}(\phi_1+i\gamma_5\phi_2)\Psi +\frac{g_2}{24}(\phi_i\phi_i)^2]\,.
\end{equation} 
which has weakly coupled IR fixed points in $d=4-\epsilon$. 

The GN and NJL models can be seen as being a part of a more general series of models given by the action 
\begin{equation}\label{genmodel}
    S=-\int d^dx\Big[\Bar{\Psi}\slashed{\partial}\Psi+\frac{g}{2}\sum_{a=1}^{N_s}(\Bar{\Psi}\Sigma_a\Psi)^2\Big]
\end{equation}
Where $\Sigma_a$ are matrices that obey certain conditions described in section \ref{sec:genns}. The UV fixed points of such models can be again studied within the large $N$ expansion, by performing a Hubbard-Stratonovich transformation that introduces $N_s$ scalar auxiliary fields. By a similar logic to the one that applies to the GN and NJL models, one expects that the general models (\ref{genmodel}) have a UV completion in terms of a theory with $N_s$ scalars and $N_f$ fermions with action
\begin{equation}
     S_{N_f,N_s}=\int d^dx\left[\frac{1}{2}(\partial\vec{\phi})^2-\Bar{\Psi}\slashed{\partial}\Psi + +g_1\sum_a \phi_a\Bar{\Psi}\Sigma_a\Psi+ \frac{g_2}{24}(\vec{\phi}\cdot\vec{\phi})^2\right]. 
\end{equation}
Here $\vec{\phi}=\begin{pmatrix}
        \phi_1 & \phi_2 &\dots &\phi_{N_s}
        \end{pmatrix}$ denote the $N_s$ real scalar fields transforming in the fundamental of the $O(N_s)$ symmetry. 
The cases $N_s=1$ and $N_s=2$ are the GNY and NJLY models, respectively. 
The $N_s=3$ model is known as the chiral Heisenberg model,  
and it has been proposed to describe the behavior of the antiferromagnetic critical point of graphene \cite{PhysRevB.89.205403}. In the quartic formulation (\ref{genmodel}), it is also known as the $SU(2)$ Gross-Neveu model \cite{PhysRevD.97.105009}, and it can be studied in the $d=2+\epsilon$ expansion as well.

As shown in \cite{giombi2025linedefectsfermioniccfts} for the GNY model, and later generalized to NJLY and chiral Heisenberg models in \cite{Pannell_2023}, these models admit a natural line defect operator which can be introduced as 
\begin{equation}
\label{dQFT}
S=S_{N_f,N_s}+\int_\gamma d\tau \vec{h}\cdot \vec{\phi}
\end{equation}
where $\gamma$ is the contour on which the defect is defined, and $\vec{h}=(h_1,\ldots, h_{N_s})$ are defect coupling constants. The defect breaks the $O(N_s)$ symmetry down to $O(N_s-1)$, and is analogous to the pinning field defect in the scalar $O(N)$ model \cite{Cuomo_2022}. The beta function for the defect coupling was computed to two-loop order in \cite{Pannell_2023}, and we reproduce the calculation in the Appendix for completeness. In the IR, the system flows to a non-trivial fixed point where both bulk and defect couplings are tuned to their critical values. When $\gamma$ is a straight line or circle, one hence obtains a non-trivial dCFT, and some of its observables were studied in \cite{giombi2025linedefectsfermioniccfts,Pannell_2023,barrat2025linedefectcorrelatorsfermionic}. 

\begin{figure}[h!]
     \centering
     \includegraphics[width=0.5\linewidth]{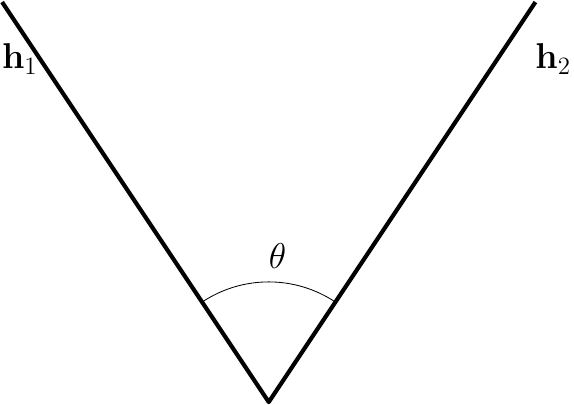}
     \caption{Cusp at angle $\theta$ with two different defect couplings on each arm.}
     \label{fig:cuspdiag}
 \end{figure}

In this paper we are interested in the case where the contour $\gamma$ in flat space is a line with a cusp as shown in figure \ref{fig:cuspdiag}. 
The defect coupling on the two arms of the cusp can be taken to be different, in general. Due to the presence of the cusp, one finds logarithmic divergences in the partition function of the theory with the line defect. We can define the corresponding cusp anomalous dimension $\Gamma_{\vec{h}_1\vec{h}_2}(\theta)$ as  \cite{cuomo2024impuritiescuspgeneraltheory}
 \begin{equation}\label{flspcadded}
\log\frac{Z_{\vec{h}_1\vec{h}_2}(\theta)}{Z_{CFT}}=-\Gamma_{\vec{h}_1\vec{h}_2}(\theta)\log\Big(\frac{L}{a}\Big)+\dots
 \end{equation}
The quantity on the left-hand-side is essentially the logarithm of the normalized expectation value of the defect. On the right-hand-side, $L$ and $a$ are respectively the IR and UV cutoffs. The ellipses indicate terms with non-logarithmic divergences.\footnote{One generally  finds power-law divergences, including ``cosmological constant" renormalization terms that are proportional to $L/a$.}

An alternative way to obtain the cusp anomalous dimension is to perform a Weyl transformation to the cylinder $\mathbb{R}_{\tau}\times S^{d-1}$.
Under this map, the radial coordinate which gives the distance from the cusp in flat space is mapped to the Euclidean time direction $\mathbb{R}_{\tau}$. 
The cusped contour on $\mathbb{R}^d$ thus maps in the cylinder frame to two lines running along $\mathbb{R}_{\tau}$ and separated by an angle $\theta$ on $S^{d-1}$. In this picture, the cusp anomalous dimension is mapped to the ground state energy of the system in the presence of the two defect lines. In order to unambigously define the ground state energy, one has to subtract the ``self-energy" contributions due to each line \cite{cuomo2024impuritiescuspgeneraltheory}
\begin{equation}\label{eq:caddef}
    \log \frac{Z_{\vec{h}_1\vec{h}_2}(\theta)}{\sqrt{Z_{\vec{h}_1\vec{h}_1}(\pi)Z_{\vec{h}_2\vec{h}_2}(\pi)}}=-\Gamma_{\vec{h}_1\vec{h}_2}(\theta)T
\end{equation}
where $Z_{\vec{h_1}\vec{h_2}}(\theta)$ is the partition function on $\mathbb{R}_{\tau}\times S^{d-1}$ computed in the presence of the two lines separated by the angle $\theta$. On the right-hand-side, $T$ denotes the length of the Euclidean time axis, which appears as an overall factor due to the time-translational invariance on the cylinder. Note that an infinite line in flat space with no cusp is mapped to two lines on the cylinder separated by an angle $\theta=\pi$, which explains the form of the normalization factors on the left-hand-side. Indeed, note that the expression above is appropriately normalized to give $\Gamma_{\vec{h}\vec{h}}(\pi)=0$, since an infinite line with no cusp and with uniform coupling along the line has no logarithmic divergences. 

The cusp anomalous dimension encodes several interesting observables of the dCFT \cite{cuomo2024impuritiescuspgeneraltheory}. For example, when $\theta=\pi$ and for unequal couplings on the two arms of the cusp, it is directly related to the dimension of the so-called defect changing operator (which can be thought as a local operator $O_{\vec{h}_1\vec{h}_2}$ inserted at the junction between the two half lines) as 
\begin{equation}\label{eq:cadtodcodim}
    \Gamma_{\vec{h}_1\vec{h}_2}(\pi)=\Delta_{\vec{h}_1\vec{h}_2}
\end{equation}
Similarly, one can extract the dimension of the defect creation (annihilation) operator by setting $\vec{h}_2=0$ in (\ref{eq:cadtodcodim}): This describes a trivial defect changing into a non-trivial one, or vice-versa. The dimension of the defect changing operators play an important role in the question of stability of the line defects that spontaneously break the global symmetry \cite{Lanzetta:2025xfw}.

\begin{figure}[h!]
    \centering
    \includegraphics[width=0.5\linewidth]{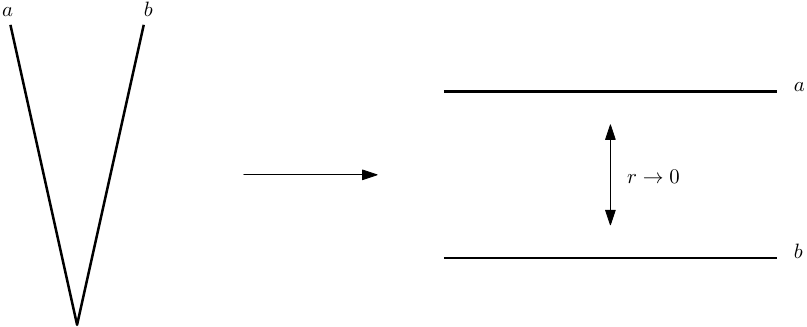}
    \caption{The two defects $a$ and $b$ with cusp angle $\theta\rightarrow 0$ can be described as two line defects $a$ and $\bar{b}$ fusing.}
    \label{fig:fusdiag}
\end{figure}

An interesting regime of the cusp anomalous dimension is the $\theta\rightarrow 0$ limit, which can be used to extract information on the fusion of two line defects, as shown in fig. \ref{fig:fusdiag}. 
As explained in \cite{cuomo2024impuritiescuspgeneraltheory}, for small $\theta$ one expects the cusp anomalous dimension to have the structure
\begin{equation}\label{casimirdefcad}
\Gamma_{\vec{h}_1 \vec{h}_2}(\theta\rightarrow0)=\frac{C_{\vec{h}_1 \vec{h}_2 \vec{h}_{\rm fus}}}{\theta}+\Delta_{\vec{h}_{\rm fus}0} + \alpha \theta^{\Delta_{\rm irr}-1}+\ldots
\end{equation}
where $\Delta_{\vec{h}_{\rm fus}0}$ is the defect creation operator corresponding to the fused coupling $\vec{h}_{\rm fus}$, $\Delta_{\rm irr}$ is the dimension of the least irrelevant operator on the defect, and the fusion coefficients $C_{\vec{h}_1 \vec{h}_2 \vec{h}_{\rm fus}}$ are related to the Casimir energy between the two defect lines separated by a distance $r\sim \theta$. After computing the cusp anomalous dimension, we will examine the small $\theta$ limit in section \ref{sec:casimir}, confirming the structure (\ref{casimirdefcad}). For the GNY model, we find the corresponding Casimir energies to be in agreement with those previously computed in \cite{diatlyk2024defectfusioncasimirenergy}. 
 

In the opposite limit of a near straight line, $\theta \rightarrow \pi$, it is not difficult to see that the cusp anomalous dimension encodes integrated correlation functions of the displacement operator, which arises from breaking of translational invariance by the defect and has protected dimension $\Delta=2$. In particular, as we review in section \ref{disptiltsec}, the quadratic term in the expansion of  $\Gamma_{\vec{h}\vec{h}}(\theta)$ near $\theta=\pi$ is related to the displacement 2-point function $\langle D(t_1)D(t_2)\rangle = C_D/(t_1-t_2)^4$ integrated over the defect, and it can be used to extract the $C_D$ coefficient \cite{Correa:2012at, Cavagli__2023, cuomo2024impuritiescuspgeneraltheory}.\footnote{More generally, for unequal couplings the expansion of the cusp anomalous dimension near $\theta=\pi$ has the structure $\Gamma_{\vec{h}_1\vec{h}_2}(\theta)=\Delta_{\vec{h}_1\vec{h}_2}+\frac{1}{2}\beta(\theta-\pi)^2+\ldots$, with $\beta$ related to the integrated 4-point function of two displacement and two defect changing operators \cite{cuomo2024impuritiescuspgeneraltheory}.} 

Due to the breaking by the defect (\ref{dQFT}) of the $O(N_s)$ global symmetry to $O(N_s-1)$, one also has (for $N_s>1$) another set of protected operators with $\Delta=1$, known as the tilt operators. Let us introduce an angle in the scalar coupling space as $\vec{h}_1\cdot\vec{h}_2=|\vec{h}_1||\vec{h}_2|\cos\theta_f$, so that we can write $\Gamma_{\vec{h}_1,\vec{h}_2}(\theta)\equiv \Gamma(\theta, \theta_f)$. One can then see that the expansion of $\Gamma(\theta, \theta_f)$ near $\theta_f=0$ and $\theta=\pi$ encodes integrated correlators of tilt and displacement operators. For instance, the quadratic term in $\theta_f$ at $\theta=\pi$ encodes the normalization coefficient $C_t$ appearing in the tilt two-point function \cite{Correa:2012at, Cavagli__2023, cuomo2024impuritiescuspgeneraltheory}. In section \ref{disptiltsec}, we extract both $C_D$ and $C_t$ from our result for the cusp anomalous dimension, finding agreement with results previously obtained by a direct calculation \cite{barrat2025linedefectcorrelatorsfermionic}.

The rest of this paper is organized as follows. In section \ref{sec:cusp-calc} we compute the cusp anomalous dimension to one-loop order in the epsilon expansion, starting from the GNY model and then generalizing to the theories (\ref{dQFT}) with $N_s$ scalar fields. In section \ref{sec:results} we discuss several observables that can be extracted from the cusp anomalous dimension, and obtain some estimates of their values in $d=3$ by Pad\'e extrapolations of the one-loop results. Some technical details of the calculation of the cusp anomalous dimension are collected in appendix \ref{app:appA}, while the two-loop beta functions and fixed point values for bulk and defect couplings are summarized in appendix \ref{app:appB}.

\section{Calculating the Cusp Anomalous Dimension}
\label{sec:cusp-calc}
We start by describing the calculation of the cusp anomalous dimension in the GNY model. 
It will then be straightforward to generalize the calculation to the models with additional scalar fields.

\subsection{Cusp Anomalous dimension in the GNY model}\label{sec:cadgny}
The action for the GNY model in the presence of the defect is given by
\begin{equation}
S = S_{\rm GNY}+\int_{\gamma} d\tau h \phi
\end{equation}
with the bulk action given in (\ref{GNY}). Below we will denote $N=N_f c_d$, with $c_d$ being the number of components of a Dirac fermion, which we take to be fixed to $c_d=4$ (the components of a 4d Dirac fermion). 

In flat space, the scalar and fermion propagators in our conventions are given by
\begin{equation}
    G_{flat}^{scalar}(x,y)=\langle s(y)s(x)\rangle=\mathcal{N}_{d}\frac{1}{|x-y|^{d-2}}
\end{equation}
\begin{equation}
G_{flat}^{fermion}(x,y)=\langle\Bar{\Psi}(y)\Psi(x)\rangle=\mathcal{N}_d(d-2)\gamma^\mu(y_\mu-x_\mu)\frac{1}{|x-y|^{d}}
\end{equation}  
Here, we have defined
\begin{equation}\label{eq:Nddef}
    \mathcal{N}_d=\frac{\Gamma\big(\frac{d}{2}-1\big)}{4\pi^\frac{d}{2}}\,.
\end{equation}

As explained in the introduction, to compute the cusp anomalous dimension it is convenient to go to the Weyl cylinder frame. 
 Writing the flat space metric as
  \begin{equation}
  ds_{flat}^2=dr^2+r^2d\Omega_{d-1}^2 \,,
  \end{equation}
we may perform a Weyl transformation by noting that
  \begin{equation}
      ds_{flat}^2=\frac{r^2}{R^2}ds_{cyl}^2\,,
  \end{equation}
where, after defining the Euclidean time variable $\tau$ as $\tau=\log \frac{r}{R}$, the cylinder metric is given by 
\begin{equation}
    ds_{cyl}^2=d\tau^2+R^2d\Omega_{d-1}^2\,.
\end{equation}
For the calculation below, we will set $R=1$ for convenience. The scalar and fermion propagators in the Weyl cylinder frame are given by (see e.g. \cite{Birrell_Davies_1982})
\begin{equation}
G_{cyl}^{scalar}(\tau_1,\hat{n}_1;\tau_2,\hat{n}_2)=\mathcal{N}_d\frac{1}{(2\cosh(\tau_1-\tau_2)-2\hat{n}_1\cdot\hat{n}_2)^{\frac{d}{2}-1}}
\end{equation}
\begin{equation}    G_{cyl}^{fermion}=(d-2)\mathcal{N}_d\frac{1)}{(2\cosh(\tau_1-\tau_2)-2\hat{n}_1\cdot\hat{n}_2)^{\frac{d-1}{2}}}\gamma^\mu u_{\mu}
\end{equation}
where $u_\mu=\frac{y-x}{|y-x|}$ with $x=(\tau_1,\hat{n}_1)$, $y=(\tau_2,\hat{n}_2)$.

\begin{figure}[h!]
    \centering
    \begin{subfigure}[b]{0.3\textwidth}
         \centering
         \includegraphics[scale=0.8]{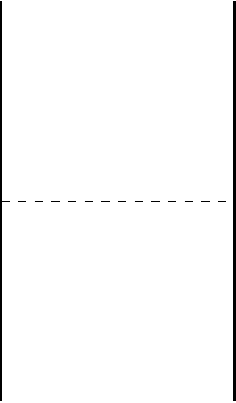}
         \caption{$\sim h_1h_2$}
         \label{linecusp}
    \end{subfigure}
    \begin{subfigure}[b]{0.3\textwidth}
         \centering
         \includegraphics[scale=0.8]{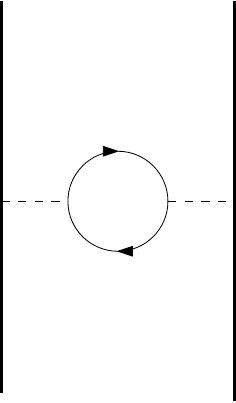}
         \caption{$\sim g_1^2Nh_1h_2$}
         \label{loopcusp}
    \end{subfigure}
    \begin{subfigure}[b]{0.3\textwidth}
         \centering
         \includegraphics[scale=0.8]{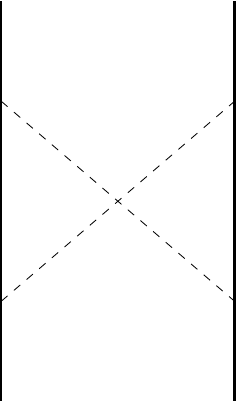}
         \caption{$\sim g_2h_1^2h_2^2$}
         \label{2L2Rcusp}
    \end{subfigure}
    \begin{subfigure}[b]{0.3\textwidth}
         \centering
         \includegraphics[scale=0.8]{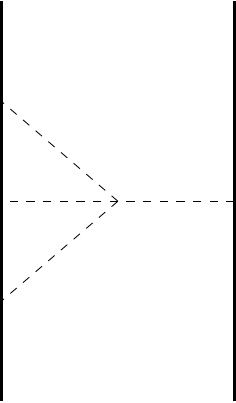}
         \caption{$\sim g_2h_1h_2^3$}
         \label{3Lcusp}
    \end{subfigure}
    \begin{subfigure}[b]{0.3\textwidth}
         \centering
         \includegraphics[scale=0.8]{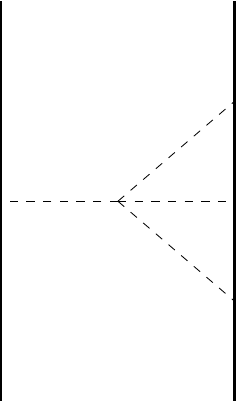}
         \caption{$\sim g_2h_1^3h_2$}
         \label{3Rcusp}
    \end{subfigure}
    \caption{Diagrams contributing to the cusp anomalous dimension to $\mathcal{O}(\epsilon)$.}
    \label{fig:cuspdiagrams}
\end{figure}

To obtain the cusp anomalous dimension from the cylinder setup, we need to compute the partition function for the theory on $R_{\tau}\times S^{d-1}$ with two defect lines along $\tau$ separated by an angle $\theta$ on the sphere (corresponding to two half-infinite lines in flat space meeting at a cusp angle $\theta$). We let $h_1$ and $h_2$ be the defect coupling strengths on the two arms of the cusp, and denote the corresponding partition function by $Z_{h_1h_2}$. Let us recall that, because of the translational invariance along $\tau$ (which reflects the scale invariance of the cusp in flat space), the calculation of all diagrams contributing to the free energy $\log Z_{h_1h_2}$ on the cylinder will include an overall constant factor  
\begin{equation}
    T\equiv \int_{-\infty}^{\infty}d\tau\,,
\end{equation}
as shown in eq. (\ref{eq:caddef}).

Working to linear order in $\epsilon$, we need to consider the five diagrams shown in figure \ref{fig:cuspdiagrams}. 
The diagrams involving just the scalar field were already computed in \cite{cuomo2024impuritiescuspgeneraltheory} in the case of the Wilson-Fisher fixed point of the $O(N)$ model, and we can borrow their results directly. The only additional diagram to this order is shown in figure \ref{loopcusp},  whose detailed calculation is given in appendix \ref{app:appA1}. 

Following \cite{cuomo2024impuritiescuspgeneraltheory}, we define
\begin{equation}\label{eq:fdef}
    f_d(x)=\int d\tau \frac{\exp\big({-(\frac{d}{2}-1)|\tau|}\big)}{\big(1+\exp(-2|\tau|)-2x\exp(-|\tau|)\big)^{\frac{d}{2}-1}}
\end{equation}
It is straightforward to see that this is the function that appears in the calculation of the free field exchange diagram in figure \ref{linecusp}, with $x=\cos\theta$. Note that in $d=4$ we have
\begin{equation}
f_{4}(\cos\theta)=\frac{\pi-\theta}{\sin\theta}
\end{equation}
To write the contribution of the diagrams in figures \ref{3Lcusp}, \ref{3Rcusp} and \ref{2L2Rcusp}, it is convenient to introduce the notations \cite{cuomo2024impuritiescuspgeneraltheory} 
\begin{equation}
    I_{11}(\hat{n}_1\cdot\hat{n}_2)=\int d^{3}\hat{n}\Big[f_4^3(\hat{n}\cdot\hat{n}_1)f_4(\hat{n}\cdot\hat{n}_2)-\frac{\pi^3f_4(\hat{n}_1\cdot\hat{n}_2)}{(2-2\hat{n}\cdot\hat{n}_1)^{\frac{3}{2}}}\Big]
\end{equation}
\begin{equation}
    I_{12}(\hat{n}_1\cdot\hat{n}_2)=\int d^3\hat{n}f_{4}^2(\hat{n}\cdot\hat{n}_1)f_4^2(\hat{n}\cdot\hat{n}_2)
\end{equation}
\begin{equation}\label{ftexp}
\Tilde{f}_d(x)=\frac{\sqrt{\pi}2^{\frac{3-d}{2}}\Gamma(\frac{d-3}{2})}{\Gamma(\frac{d-2}{2})(1-x)^{\frac{d-3}{2}}}
\end{equation}
It is not difficult to see that $I_{11}$ arises from the diagrams in figures \ref{3Lcusp} and \ref{3Rcusp}, while $I_{12}$ from the diagram in figure \ref{2L2Rcusp}.

\begin{figure}[h!]
    \centering
    \begin{subfigure}[b]{0.3\textwidth}
         \centering
         \includegraphics[scale=0.8]{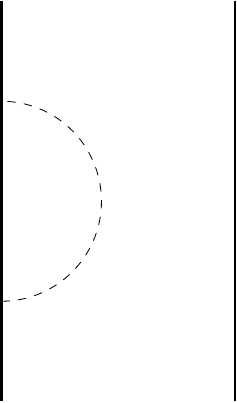}
         \caption{$\sim h_2^2$}
         \label{lineLsi}
    \end{subfigure}
    \begin{subfigure}[b]{0.3\textwidth}
         \centering
         \includegraphics[scale=0.8]{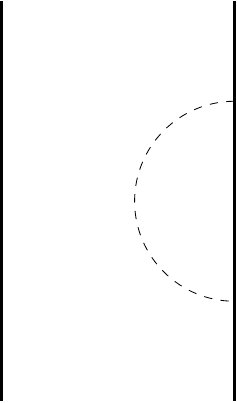}
         \caption{$\sim h_1^2$}
         \label{LineRsi}
    \end{subfigure}
    \begin{subfigure}[b]{0.3\textwidth}
         \centering
         \includegraphics[scale=0.8]{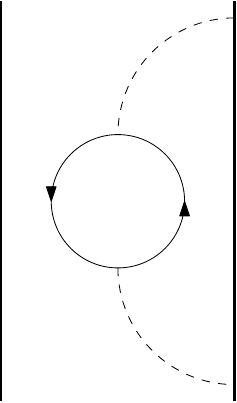}
         \caption{$\sim g_1^2h_2^2$}
         \label{loopRsi}
    \end{subfigure}
    \begin{subfigure}[b]{0.3\textwidth}
         \centering
         \includegraphics[scale=0.8]{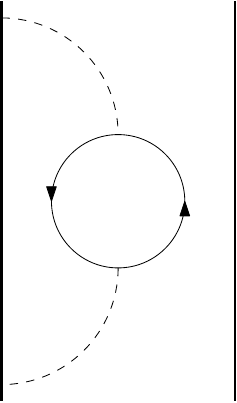}
         \caption{$\sim g_1^2h_1^2$}
         \label{loopLsi}
    \end{subfigure}
    \begin{subfigure}[b]{0.3\textwidth}
         \centering
         \includegraphics[scale=0.8]{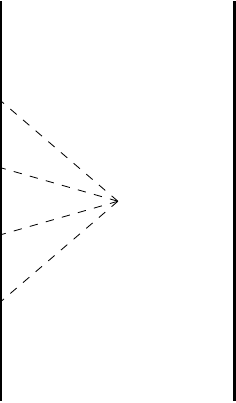}
         \caption{$\sim g_2h_2^4$}
         \label{fpLsi}
    \end{subfigure}
    \begin{subfigure}[b]{0.3\textwidth}
         \centering
         \includegraphics[scale=0.8]{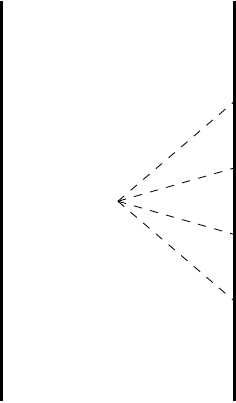}
         \caption{$\sim g_2h_1^4$}
         \label{fpRsi}
    \end{subfigure}
    \caption{Self-energy contributions to $\mathcal{O}(\epsilon)$.}
    \label{fig:sidiagrams}
\end{figure}

 Using the notations introduced above, the value of $\frac{\log Z_{h_1h_2}}{T}$ in terms of bare coupling constants (which we denote by `$0$' subscripts) is found to be
\begin{equation}\label{eq:plainpartition}
\begin{split}
    -\frac{\log Z_{h_1h_2}}{T}&=-\mathcal{N}_dh_{1,0}h_{2,0}f_{d}(\cos\theta) \\&+\frac{g_{1,0}^2Nh_{1,0}h_{2,0}}{64\pi^d}\frac{\Gamma(\frac{d}{2}-1)^2 \Gamma(d-3)}{\Gamma(d-2) (4-d)}\int d\tau \frac{\exp(-(d-3)|\tau|)}{(1-2\hat{n}_1\cdot\hat{n}_2\exp(-|\tau|)+\exp(-2|\tau|))^{d-3}} \\& +g_{2,0}h_{1,0}h_{2,0}(h_{1,0}^2+h_{2,0}^2)\frac{\mathcal{N}_d^4}{6}\left(I_{11}(\cos\theta)
+f_d(\cos\theta)\int d^{d-1}\hat{n}\Tilde{f}_d^3(\hat{n}\cdot\hat{n}_1)\right)
\\&    +g_{2,0}h_{1,0}^2h_{2,0}^2\frac{\mathcal{N}_4^4}{4}I_{12}(\cos\theta)\\&+\dots
\end{split}
\end{equation}
The term in the first line comes from the tree-level diagram in figure \ref{linecusp}. 
The term in the second line comes from the diagram in figure \ref{loopcusp}, which is computed in detail in appendix \ref{app:appA1}. Note that its expression written above is an approximate which is valid when we consider terms of $\mathcal{O}((4-d)^2)$. The term on the third line comes from the diagrams in figure \ref{3Lcusp} and figure \ref{3Rcusp}, while the term on the fourth line comes from figure \ref{2L2Rcusp}. Finally, the ellipses include the self-interaction diagrams shown in figure \ref{fig:sidiagrams}. These terms cancel out when we compute the cusp anomalous dimension as given in (\ref{eq:caddef}).

We work in the minimal subtraction scheme, where we express the bare coupling constants in terms of the physical ones through a series of poles at $\epsilon=0$. For the bulk couplings, the relation between bare and renormalized couplings takes the form
\begin{equation}\label{bulccplexp}
    g_{1,0} = M^{\frac{\epsilon}{2}} \left( g_1 + \dots \right), \hspace{0.5cm} g_{2,0} = M^{\epsilon} \left( g_2 + \dots \right)
\end{equation}
with $M$ the renormalization scale. For the defect coupling, using the results in \cite{giombi2025linedefectsfermioniccfts} we have  
\begin{equation}\label{eq:hctexp}
h_0 = M^{\frac{\epsilon}{2}}h \left( 1 + \frac{\delta h}{\epsilon} + \dots \right). 
\end{equation}
\begin{equation}\label{eq:hct}
\delta h = \frac{g_2 h^3}{192 \pi^2 } + \frac{g_1^2 h N}{32 \pi^2 } + \dots.  
\end{equation}

Expressing (\ref{eq:plainpartition}) in terms of the renormalized couplings, we have
\begin{equation}\label{eq: partitionwcounterterms}
    \begin{split}
        -\frac{\log Z_{h_1h_2}}{T}&=-M^{\epsilon}\mathcal{N}_dh_1h_2f_{d}(\cos\theta)(1+\frac{\delta h_1}{\epsilon}+\frac{\delta h_2}{\epsilon}) \\&+ M^{2\epsilon}f_d(\cos\theta)g_2h_1h_2(h_1^2+h_2^2)\frac{\mathcal{N}_d^4}{6}\int d^{d-1}\hat{n}\Tilde{f}_d^3(\hat{n}\cdot\hat{n}_1)\\&+M^{2\epsilon}\frac{g_1^2Nh_1h_2}{64\pi^d}\frac{\Gamma(\frac{d}{2}-1)^2 \Gamma(d-3)}{\Gamma(d-2) (4-d)}\int d\tau \frac{\exp(-(d-3)|\tau|)}{(1-2\hat{n}_1\cdot\hat{n}_2\exp(-|\tau|)+\exp(-2|\tau|))^{d-3}} \\& +g_2h_1h_2(h_1^2+h_2^2)\frac{\mathcal{N}_4^4}{6}I_{11}(\cos\theta)\\&
    +g_2h_1^2h_2^2\frac{\mathcal{N}_4^4}{4}I_{12}(\cos\theta)\\&+\dots
    \end{split}
\end{equation}
The integral of $\Tilde{f}^3$ can be easily evaluated as
\begin{equation}
    \int d^{d-1}\hat{n}\Tilde{f}_d^3(\hat{n}\cdot\hat{n}_1)=\frac{2\pi^{\frac{d-1}{2}}}{\Gamma(\frac{d-1}{2})}\int_0^\pi d\theta \sin ^{d-2}\theta \Tilde{f}_d^3(\cos\theta)\,.
\end{equation}
Substituting the expression (\ref{ftexp}), we obtain
\begin{equation}\label{ftcalc}
    \int d^{d-1}\hat{n}\Tilde{f}_d^3(\hat{n}\cdot\hat{n}_1)=\frac{\pi^{\frac{d+2}{2}}2^{8-2d}\Gamma(\frac{d-3}{2})^3\Gamma(4-d)}{\Gamma(\frac{7-d}{2})\Gamma(\frac{d-2}{2})^3}
\end{equation}
Let us also introduce for convenience the notation
\begin{equation}\label{eq:Fdef}
    F_{\epsilon}(x)=\int d\tau \frac{ \exp(-(1-\epsilon)|\tau|)}{(1-2x\exp(-|\tau|)+\exp(-2|\tau|))^{1-\epsilon}}
\end{equation}
which is the function appearing in the fermion loop diagram. Note that 
\begin{equation}\label{eq:fFrel}
f_{4-\epsilon}(x)=F_{\frac{\epsilon}{2}}(x)
\end{equation}
We will also need its derivative at $\epsilon=0$, which we denote as
\begin{equation}\label{eq:F'def}
    F'_0(x)=\frac{dF_{\epsilon}}{d\epsilon} \Big|_{ \epsilon=0}
\end{equation}

Then, substituting (\ref{ftcalc}) in (\ref{eq: partitionwcounterterms}), using $d=4-\epsilon$, and keeping terms up to order $\epsilon$, we get
 \begin{equation}\label{eq:partitionepsexp}
   \begin{split}
       \frac{\log Z_{h_1h_2}}{T}=& (1+2\epsilon\log M)\frac{h_{1}h_{2}}{4\pi^2}(F_{0}(\cos\theta)+\frac{\epsilon}{2}F'_0(\cos\theta))\times\\& \Big(1 + \frac{g_1^2N}{16\pi^2\epsilon}+\frac{g_1^2N}{32\pi^2}\log(\pi e^\gamma)-\frac{g_1^2N}{16\pi^2}\log M\\&+\frac{\epsilon}{2}\log(\pi e^\gamma M^{-2}) -\frac{g_2(h_{1*}^2+h_{2*}^2)}{192\pi^2}\log(64\pi e^{\gamma-1}M^2)\Big)\\& -(1+2\epsilon\log M)\frac{g_1^2Nh_1h_2}{64\pi^4}\left(\frac{1}{\epsilon}+\log(\pi e^{\gamma+1})\right)(F_0(\cos\theta)+\epsilon F'_0(\cos\theta))\\& -g_2h_1h_2(h_1^2+h_2^2)\frac{\mathcal{N}_4^4}{6}I_{11}(\cos\theta)\\&
    -g_2h_1^2h_2^2\frac{\mathcal{N}_4^4}{4}I_{12}(\cos\theta)\\&+\dots
   \end{split}  
 \end{equation}
Here once again, the ellipses indicate the self interaction terms which will cancel out after normalization. 
We see that, as expected, the terms with poles at $\epsilon=0$ cancel out, and we are left with
 \begin{equation}\label{eq:partitiongrouping}
     \begin{split}
         \frac{\log Z_{h_1h_2}}{T}=& \frac{h_1h_2}{4\pi^2}F_0(\cos\theta)\Big(1+\frac{\epsilon}{2}\log(\pi e^\gamma M^2 )\\& -\frac{g_1^2N}{32\pi^2}\log (\pi e^{\gamma+2}M^2)-\frac{g_2(h_1^2+h_2^2)}{192\pi^2}\log (64\pi e^{\gamma-1}M^2)\Big)\\&+\epsilon\frac{h_1h_2}{8\pi^2}F'_0(\cos\theta)-\frac{g_1^2Nh_1h_2}{128\pi^4}F'_0(\cos\theta)\\&-g_2h_1h_2(h_1^2+h_2^2)\frac{\mathcal{N}_4^4}{6}I_{11}(\cos\theta)\\&
    -g_2h_1^2h_2^2\frac{\mathcal{N}_4^4}{4}I_{12}(\cos\theta)\\&+\dots
     \end{split}
 \end{equation}
Note that the coefficient of $\log(\pi e^{\gamma} M^2)$ in this expression is proportional to $\beta_{h_1}/h_1 + \beta_{h_2}/h_2$, where $\beta_h$ is the beta function of the defect coupling given in eq. (\ref{eq:betafcn-h}). Hence, at the fixed point the dependence on $M$ drops out as expected. 

Finally, using (\ref{eq:caddef}), we obtain the cusp anomalous dimension at the fixed point\footnote{The terms we brushed under the carpet using the ellipses in (\ref{eq:partitiongrouping}) cancel fully when we normalize as per (\ref{eq:caddef}). However, there are terms in $\frac{1}{2}\Big(\log Z_{h_1h_1}(\pi)+\log Z_{h_2h_2}(\pi)\Big)$, namely the terms coming from diagrams connecting the two arms of the defect, that are left over and show up in the expression for the cusp anomalous dimension (\ref{eq:cadgen}) (these can be identified using as the terms independent of $\theta$ in the expression)}
\begin{equation}\label{eq:cadgen}
    \begin{split}
        \Gamma_{h_1h_2}(\theta) =& -\frac{1}{4\pi^2}\left(h_{1*}h_{2*}\frac{\pi-\theta}{\sin\theta}-\frac{(h_{1*}^2+h_{2*}^2)}{2}\right) \\& +\frac{g_{1*}^2N}{64\pi^4}\left(h_{1*}h_{2*}\frac{\pi-\theta}{\sin\theta}-\frac{(h_{1*}^2+h_{2*}^2)}{2}\right)\\&
+\frac{g_{2*}}{384\pi^4}\log(64e^{-1})\left(h_{1*}h_{2*}\frac{(h_{1*}^2+h_{2*}^2)}{2}\frac{\pi-\theta}{\sin\theta}
-\frac{(h_{1*}^4+h_{2*}^4)}{2}\right)\\&        -\left(\epsilon\frac{1}{8\pi^2}-\frac{g_{1*}^2N}{128\pi^4}\right)(h_{1*}h_{2*}F'_0(\cos\theta)-(h_{1*}^2+h_{2*}^2))\\&        +\frac{g_{2*}}{1536\pi^8}((h_{1*}^3h_{2*}+h_{1*}h_{2*}^3)I_{11}(\cos\theta)-(8\pi^4-4\pi^4\log8-24\pi^2\zeta(3))(h_{1*}^4+h_{2*}^4))\\&        +\frac{g_{2*}}{1024\pi^8}((h_{1*}^2h_{2*}^2)I_{12}(\cos\theta)-12\pi^2\zeta(3)(h_{1*}^4+h_{2*}^4))
    \end{split}
\end{equation}

For the GNY model, we have $h\in\mathbb{R}$ and thus, the defect couplings on either side of the cusp can differ at most by a sign.  
Hence, letting $|h_1|=|h_2|=h_*$, we can write
\begin{equation}\label{eq:cad}
    \begin{split}
        \Gamma_{h_1h_2}(\theta) =& -\frac{h_*^2}{4\pi^2}\left(\text{sgn}(h_1h_2)\frac{\pi-\theta}{\sin\theta}-1\right) \\& +\frac{h_*^2}{64\pi^4}\left(\text{sgn}(h_1h_2)\frac{\pi-\theta}{\sin\theta}-1\right)
\left(g_{1*}^2N+\frac{g_{2*}h_*^2}{6}\log(64e^{-1})\right)\\&        -\left(\epsilon\frac{h_*^2}{8\pi^2}-\frac{g_{1*}^2Nh_*^2}{128\pi^4}\right)(\text{sgn}(h_1h_2)F'_0(\cos\theta)-2)\\&       +\frac{g_{2*}h_*^4}{768\pi^8}(\text{sgn}(h_1h_2)I_{11}(\cos\theta)-8\pi^4+4\pi^4\log8+24\pi^2\zeta(3))\\&
        +\frac{g_{2*}h_*^4}{1024\pi^8}(I_{12}(\cos\theta)-24\pi^2\zeta(3))
    \end{split}
\end{equation}
Where
\begin{equation}
    \text{sgn}(x)=\begin{cases}
        -1,&\text{if}\quad x<0\\
        0,&\text{if}\quad x=0\\
        1,&\text{if}\quad x>0
    \end{cases}
\end{equation}
Plugging in the explicit expressions for the bulk and defect couplings at the fixed point, this expression gives us the explicit expression for the cusp anomalous dimension to order $\epsilon$. We discuss this, as well as various observables that can be extracted from, in section \ref{sec:results} below. 

\subsection{Generalizing to $N_s$ scalar fields}\label{sec:genns}
The calculation described in the previous section can be easily extended to the general models where we have $N_f=\frac{N}{4}$ (we take $c_d=4$ corresponding to Dirac fermion in 4d) fermion flavors, and $N_s$ scalar flavors \cite{barrat2025linedefectcorrelatorsfermionic}. 
Define
\begin{equation}\label{phidef}
    \vec{\phi}=\begin{pmatrix}
        \phi_1 & \phi_2 &\dots &\phi_{N_s}
    \end{pmatrix}
\end{equation}
\begin{equation}\label{psidef}
    \Psi=\begin{pmatrix}
        \psi_1 & \dots & \psi_{N_f}
    \end{pmatrix}
\end{equation}
where each $\phi_i$ is a real scalar field, and each $\psi_i$ is a Dirac spinor field. The action of the model in the presence of the line defect is given by
\begin{equation}\label{Sdef}
    S=S_{N,N_s}+\int_\gamma d\tau \vec{h}\cdot\vec{\phi}
\end{equation}
where $\gamma$ denotes the contour on which the defect is supported, the defect coupling $\vec{h}$ is a $N_s$-dimensional vector, and the bulk action is 
\begin{equation}\label{Sfull}
    S_{N,N_s}=\int d^dx\left[\frac{1}{2}(\partial\vec{\phi})^2-\Bar{\Psi}\slashed{\partial}\Psi + \frac{g_2}{24}(\vec{\phi}\cdot\vec{\phi})^2+g_1\sum_a \phi_a\Bar{\Psi}\Sigma_a\Psi\right]
\end{equation}
Here, we further assume that the matrices that enter the Yukawa couplings satisfy
\begin{equation}\label{trcond}
{\rm tr}_{flavor}{\rm tr}_{spinor}(\Sigma_a\gamma^\mu\Sigma_b\gamma^\nu)=N\eta^{\mu\nu}\delta_{ab} \quad \forall a,b=1,\dots,N_s
\end{equation}
where $\gamma^\mu$ are matrices satisfying the Clifford algebra in $d$ dimensions. Furthermore, for the purpose of our calculation, we also assume the $\Sigma_a$ to additionally obey
\begin{equation}
\gamma^\mu\Sigma_a\gamma^\nu\Sigma_b\gamma^\lambda=(2\delta_{ab}-1)\gamma^\mu\gamma^\nu\gamma^\lambda \quad \forall a,b=1,\dots,N_s
\end{equation}
While these two conditions restrict the allowed $\Sigma_a$, they are sufficient to describe the interesting GNY, NJLY and chiral Heisenberg models. 
The GNY model is the simplest possibility, corresponding to a single scalar and $N_f$ fermions coupled via the identity matrix. Namely, we have $N_s=1$ and $\Sigma_1=\mathbb{1}_{N_f\times N_f}\otimes\mathbb{1}_{spinor}$. The next simplest model is the NJLY model with two scalar fields. We have $N_s=2$, and the corresponding coupling matrices are $\Sigma_1=\mathbb{1}_{N_f\times N_f}\otimes\mathbb{1}_{spinor}$ and $\Sigma_2=\mathbb{1}_{N_f\times N_f}\otimes i\gamma^5$. The third model we investigate is the chiral Heisenberg model. In this model, we have the fermions transforming under a $2N_f\times 2N_f$ dimensional representation of the Clifford algebra \cite{barrat2025linedefectcorrelatorsfermionic}, with
\begin{equation}
    \begin{split}
        \Psi&=(\Psi_+,\Psi_-)\\
    \Psi_{\pm}&=(\psi^1_\pm,\dots,\psi^{N_f}_\pm)\\
    \slashed{\partial}&=(\mathbb{1}_{2\times2}\otimes\gamma^\mu)\partial_\mu\\
    \Sigma_a&=\mathbb{1}_{2N_f\times2N_f}\otimes\sigma^a
    \end{split}
\end{equation}
Here, the $\sigma^a$ are Pauli matrices normalized to obey $\sigma^a\sigma^a=1$.

In the general model with $N_s$ scalars, we may parametrize the defect coupling $h\in\mathbb{R}^{N_s}$ as
\begin{equation}\label{defcoupspec}
    \vec{h}=h \hat{m}
\end{equation}
Here, $h$ is the magnitude of the fixed point defect coupling, while $\hat{m}$ is a real unit vector on $S^{N_s-1}$. 
The beta function for the defect coupling up to 2-loop order is given by (see appendix \ref{app:appB} and \cite{barrat2025linedefectcorrelatorsfermionic, Pannell_2023})\footnote{There appears to be a typo in one of the terms of the beta function given in \cite{barrat2025linedefectcorrelatorsfermionic}: an additional factor of $\frac{1}{2}$ in the term that goes as $g_1^4h$.}
\begin{align}\label{hbetafuncmain}
    \beta_h =& -\frac{\epsilon h}{2} + \frac{1}{(4\pi)^2} \left(\frac{g_1^2Nh}{2}+\frac{g_2 h^3}{6}\right) + \frac{1}{(4\pi)^4} \Bigg\{g_2^2 h \left(\frac{(2 + N_s)}{36} - \frac{h^2 (N_s+8)}{36} - \frac{h^4}{12}\right) - \frac{g_2 g_1^2 h^3 N}{4} \notag \\
 & + g^4_1 h \left(- \frac{(N_s+4)N}{4} + h^2 N \left(1 - \frac{\pi^2}{6}\right)\right)\Bigg\}
\end{align}
From the beta function, we see that, in addition to the trivial UV fixed point $h=0$, there is a non-trivial IR fixed point of the form 
\begin{equation}\label{hform}
    h_*^2=h_{*,0}^2+\epsilon h_{*,1}^2 + \mathcal{O}(\epsilon^2)
\end{equation}
where
\begin{equation}\label{h0eqsec2}
    h_{*,0}^2=\frac{4(4-N_s)(N_s+8)}{\Big( 8 - N - 2 N_s + \sqrt{ N^2 + 4 (-4 + N_s)^2 + 4 N (28 + 5 N_s) } \Big)}
\end{equation}
The expression for $h_{*,1}^2$ is very large and messy, and it is given in appendix \ref{app:appB} along with other details of the calculation.

From the beta function, one can also read off the dimension of the lowest defect scalar $\hat{\phi}$ as
\begin{equation}
\label{eq:beta-der}
    \Delta_{\hat{\phi}}-1=\frac{\partial\beta_h}{\partial h}\Bigg|_{h=h_*}
\end{equation}
When applied at the trivial fixed point $h_*$, this implies that the linear term in the beta function (\ref{hbetafuncmain}) should give us the dimension of the bulk operator $\phi$. Indeed, we find  
\begin{equation}
\Delta_\phi =1+ \frac{\partial\beta_h}{\partial h}\Bigg|_{h=0}=1-\frac{\epsilon}{2}+\frac{g_{1*}^2N}{2(4\pi)^2}+\frac{g_{2*}^2(2+N_s)}{36(4\pi)^4}-\frac{g_{1*}^4N(4+N_s)}{4(4\pi)^4}
\end{equation}
This matches the known results for the (bulk) scalar anomalous dimension in these models \cite{MACHACEK198383,MACHACEK198570,Machacek:1983fi,Moshe_2003,VANDAMME1984105,ROSENSTEIN1993381}.

At the non-trivial fixed point, using (\ref{eq:beta-der}) we obtain the dimension of the defect scalar operator $\hat{\phi}$ to be
\begin{align}\label{dphihat}
    \Delta_{\hat{\phi}}=& 1-\frac{\epsilon }{2} + \frac{1}{(4\pi)^2} \left(\frac{g_{1*}^2N}{2}+\frac{g_{2*} h_*^2}{2}\right) + \frac{1}{(4\pi)^4} \Bigg\{g_{2*}^2  \left(\frac{(2 + N_s)}{36} - \frac{h_*^2 (N_s+8)}{12} - \frac{5h_*^4}{12}\right) - \frac{3g_{2*} g_{1*}^2 h_*^2 N}{4} \notag \\
 & + g^4_{1*}  \left(- \frac{(N_s+4)N}{4} + 3h_*^2 N \left(1 - \frac{\pi^2}{6}\right)\right)\Bigg\}
\end{align}
Since the beta function is known to two loop order, we can obtain the explicit expression for $\Delta_{\hat{\phi}}$ to order $O(\epsilon^2)$ simply by plugging the explicit fixed point couplings into (\ref{dphihat}). Showing for brevity only terms up to order $\epsilon$, we have
\begin{equation}
    \Delta_{\hat{\phi}}=1+\epsilon\frac{2(4-N_s)}{N+8-2N_s}+O(\epsilon^2)
\end{equation}
This matches the expression obtained in \cite{barrat2025linedefectcorrelatorsfermionic}.
The direct truncation of the resulting $\epsilon$ expansion (including the $O(\epsilon^2)$ terms) appears to give a highly oscillatory series as we take $\epsilon \rightarrow 1$. The Pad\'e approximations are more stable, and we present the numerical values in table \ref{tab:data1}.

Let us now move on to the calculation of the cusp anomalous dimension. One needs to compute the same diagrams shown in figure \ref{fig:cuspdiagrams}, where now the defect couplings on each line are given by the vectors $\vec{h}_1$ and $\vec{h}_2$. All the diagrams involving only the scalar fields have the same form as the corresponding diagrams in the $O(N_s)$ model computed in \cite{cuomo2024impuritiescuspgeneraltheory}.  The one additional diagram in figure \ref{loopcusp} is clearly proportional to $\vec{h}_1\cdot \vec{h}_2$.

At the fixed point, we get the final result for the cusp anomalous dimension to be  
\begin{equation}\label{gengammaexpmixed}
    \begin{split}
        \Gamma_{\vec{h}_{1*}\vec{h}_{2*}}(\theta) =& -\frac{1}{4\pi^2}\left(\vec{h}_{1*}\cdot\vec{h}_{2*}\frac{\pi-\theta}{\sin\theta}-\frac{(\vec{h}_{1*}^2+\vec{h}_{2*}^2)}{2}\right) \\& +\frac{g_{1*}^2N}{64\pi^4}\left(\vec{h}_{1*}\cdot\vec{h}_{2*}\frac{\pi-\theta}{\sin\theta}-\frac{(\vec{h}_{1*}^2+\vec{h}_{2*}^2)}{2}\right)\\& +\frac{g_{2*}}{384\pi^4}\log(64e^{-1})\left(\vec{h}_{1*}\cdot\vec{h}_{2*}\frac{(\vec{h}_{1*}^2+\vec{h}_{2*}^2)}{2}\frac{\pi-\theta}{\sin\theta}-\frac{((\vec{h}_{1*}^2)^2+(\vec{h}_{2*}^2)^2)}{2}\right)\\&
        -\left(\epsilon\frac{1}{8\pi^2}-\frac{g_{1*}^2N}{128\pi^4}\right)((\vec{h}_{1*}\cdot\vec{h}_{2*})F'_0(\cos\theta)-(\vec{h}_{1*}^2+\vec{h}_{2*}^2))\\&
        +\frac{g_{2*}}{1536\pi^8}\Big((\vec{h}_{1*}\cdot\vec{h}_{2*})(\vec{h}_{1*}^2+\vec{h}_{2*}^2)I_{11}(\cos\theta)-(8\pi^4-4\pi^4\log8-24\pi^2\zeta(3))((\vec{h}_{1*}^2)^2+(\vec{h}_{2*}^2)^2)\Big)\\&
        +\frac{g_{2*}}{1024\pi^8}\Big(\left(\frac{\vec{h}_{1*}^2\vec{h}_{2*}^2+2(\vec{h}_{1*}\cdot\vec{h}_{2*})^2}{3}\right)I_{12}(\cos\theta)-12\pi^2\zeta(3)\big((\vec{h}_{1*}^2)^2+(\vec{h}_{2*}^2)^2\big)\Big) 
    \end{split}
\end{equation}
If we assume that both sides of the cusp have the defect coupling to be of the form (\ref{defcoupspec}), with the same magnitude $h_*$ corresponding to the non-trivial fixed point on both lines, we get the cusp anomalous dimension to be
\begin{equation}\label{gengammaexp}
    \begin{split}
        \Gamma_{\vec{h}_1\vec{h}_2}(\theta) =& -\frac{h_*^2}{4\pi^2}\left(\hat{m}_1\cdot\hat{m}_2\frac{\pi-\theta}{\sin\theta}-1\right) \\& +\frac{h_*^2}{64\pi^4}\left(\hat{m}_1\cdot\hat{m}_2\frac{\pi-\theta}{\sin\theta}-1\right)\left( g_{1*}^2N+\frac{g_{2*}h_*^2}{6}\log(64e^{-1})\right)\\&
-\left(\epsilon\frac{h_*^2}{8\pi^2}-\frac{g_{1*}^2Nh_*^2}{128\pi^4}\right)((\hat{m}_1\cdot\hat{m}_2)F'_0(\cos\theta)-2)\\&        +\frac{g_{2*}h_*^4}{768\pi^8}((\hat{m}_1\cdot\hat{m}_2)I_{11}(\cos\theta)-8\pi^4+4\pi^4\log8+24\pi^2\zeta(3))\\&
+\frac{g_{2*}h_*^4}{1024\pi^8}(\left(\frac{1+2(\hat{m}_1\cdot\hat{m}_2)^2}{3}\right)I_{12}(\cos\theta)-24\pi^2\zeta(3)) 
    \end{split}
\end{equation}
We see that if we set $N_s=1$, we recover the form expected from (\ref{eq:cad}). As a check, we can also compare this result to the one obtained in \cite{cuomo2024impuritiescuspgeneraltheory} by formally setting $N=0$, which is expected to correspond to the scalar $O(N_s)$ model in this limit.

After plugging in the explicit expressions for the fixed point couplings, the cusp anomalous dimension takes the form
\begin{equation}
\Gamma_{\hat{m}_1\hat{m}_2}(\theta)=\Gamma_0(\theta)+\epsilon\Gamma_1(\theta)+\mathcal{O}(\epsilon^2),
 \end{equation}
where, after substituting the explicit fixed point values given in appendix \ref{app:appB}, we get
\begin{equation}
    \Gamma_0(\theta)=-\frac{(4-N_s)(N_s+8)}{\pi^2\Big( 8 - N - 2 N_s + \sqrt{ N^2 + 4 (-4 + N_s)^2 + 4 N (28 + 5 N_s) } \Big)}\left(\hat{m}_1\cdot\hat{m}_2\frac{\pi-\theta}{\sin\theta}-1\right)
\end{equation}
and
\begin{equation}
    \begin{split}
        \Gamma_1(\theta)=& -\frac{h_{*,1}^2}{4\pi^2}\left(\hat{m}_1\cdot\hat{m}_2\frac{\pi-\theta}{\sin\theta}-1\right)\\&-\frac{h_{*,0}^2}{4\pi^2}\Bigg[-\frac{(N-4+N_s+(4-N_s)\log64)}{N+8-2N_s}\left(\hat{m}_1\cdot\hat{m}_2\frac{\pi-\theta}{\sin\theta}-1\right) \\&+\Big(\frac{(8-2N_s)}{2(N+8-2N_s)}(\hat{m}_1\cdot\hat{m}_2F'_0(\cos\theta)-2)\Big)\\&-\frac{3(4-N_s)}{2\pi^4(N+8-2N_s)}\times\\&\Big(\frac{1}{3}((\hat{m}_1\cdot\hat{m}_2)I_{11}(\cos\theta)-8\pi^4+4\pi^4\log8+24\pi^2\zeta(3))\\&+\frac{1}{4}(\left(\frac{1+2(\hat{m}_1\cdot\hat{m}_2)^2}{3}\right)I_{12}(\cos\theta)-24\pi^2\zeta(3)) \Big)\Bigg]
    \end{split}
\end{equation}
with $h_{*,0}$, $h_{*,1}$ defined by (\ref{hform}). We will discuss the epsilon expansion results in more detail in section \ref{sec:results} below.

\subsection{Large $N$ limit}\label{sec:largeN}
In addition to the epsilon expansion in $d=4-\epsilon$, we can also consider the large $N$ expansion at general $d$. This provides a cross-check of the results obtained in the previous sections.  

For a Yukawa theory of the form (\ref{Sdef}), we can define the corresponding theory with a quartic fermion vertex to be
\begin{equation}
    S=-\int d^{d}x\Big[\Bar{\Psi}\slashed{\partial}\Psi +\frac{g}{2}\Big(\sum_{a=1}^{N_s} (\Bar{\Psi}\Sigma_a\Psi)^2\Big)\Big]
\end{equation}
To develop the large $N$ expansion, we can perform a Hubbard-Stratanovich transformation to obtain the action in terms of auxiliary scalar fields $\{ \sigma_a \}$
 \begin{equation}
     S=-\int d^dx\Big[\Bar{\Psi}\slashed{\partial}\Psi +\sum_{a=1}^{N_s} \Big(\frac{1}{\sqrt{N}}\sigma_a\Bar{\Psi}\Sigma_a\Psi -\frac{\sigma_a^2}{2gN}\Big)\Big]
 \end{equation}
The term in the action quadratic in the auxiliary fields becomes irrelevant in the UV limit and can be dropped for the purpose of developing the $1/N$ expansion of the fixed point. Integrating out the fermion fields, one obtains the resummed propagator of the auxiliary fields in the bulk to be \cite{Goykhman_2021,ZINNJUSTIN1991105,Moshe_2003}\footnote{Note that the relation (\ref{trcond}) is necessary to claim the results below. Should the matrices not be of the form we have assumed, we get cross terms. To remove these, we will need to rotate our basis in the $O(N_s)$ space to obtain a relation equivalent to (\ref{trcond}) among the new $\Sigma_a$ matrices.}
 \begin{equation}\label{LNORTHO}
     \langle \sigma_a(x)\sigma_b(0)\rangle=\frac{\mathcal{N}_{\sigma}}{x^2}\delta_{ab}
 \end{equation}
 Where,
 \begin{equation}
     \mathcal{N}_\sigma=-\frac{2^d\sin(\frac{\pi d}{2})\Gamma(\frac{d-1}{2})}{\pi^{\frac{3}{2}}\Gamma(\frac{d}{2}-1)}
 \end{equation}
This implies that at large $N$ the auxiliary fields behave as conformal operators with scaling dimension $\Delta_{\sigma}=1+O(1/N)$.  

In this large $N$ treatment, the counterpart of the line defect (\ref{dQFT}) is obtained by coupling the line to the auxiliary field \cite{giombi2025linedefectsfermioniccfts}
 \begin{equation}
     S_{\text{defect}} = h\int_\gamma d\tau \sigma_1
 \end{equation}
 Here, $\gamma$ is the contour on which the defect is supported, and we have chosen for simplicity the basis in the defect coupling space such that the defect couples to the first auxiliary field $\sigma_1$. For the calculation of the beta function, the contour $\gamma$ will just be a straight line along the $\tau$ direction. We later promote this to a line with a cusp to calculate the cusp anomalous dimension as done earlier.

For the purpose of developing the perturbative renormalization at large $N$, one may introduce a regulator $\delta$ as
 \begin{equation}
     \langle\sigma_{a,0} (x) \sigma_{b,0}(0) \rangle = \frac{\mathcal{N}_{\sigma} }{|x|^{2 + \delta}}\delta_{ab}
 \end{equation}
 This makes the dimension of the bare $\sigma_a$ fields $1 + \delta/2$ at large $N$. Consequently, the bare defect coupling $h_0$ also acquires a dimension $-\delta/2$. 
Denoting all bare quantities with a `0' subscript, the complete action describing the dCFT at large $N$ can be written as 
\begin{equation}
S = -\int d^d x \left( \bar{\Psi}_i \slashed{\partial} \Psi^i + \frac{1}{\sqrt{N}}\sum_{a=1}^{N_s} \sigma_{a,0} \bar{\Psi}_i \Psi^i   \right) + h_0 \int d \tau \sigma_{1,0}(\tau, \mathbf{0}),
\end{equation} 
and the large $N$ perturbation theory can be developed by using the standard propagators for the fermions and the resummed propagator (\ref{LNORTHO}) for the auxiliary fields. 

The bare defect coupling can be expressed in terms of the renormalized one as
\begin{equation}
h_0 = M^{-\delta/2}h\left(1+\frac{\delta_1 h}{\delta} + \dots\right).
\end{equation}
The diagrams contributing to the renormalization of $h$ to $O(\frac{1}{N})$ come from the one point function of $\sigma_1$
\begin{equation}\label{eq:sig10exp}
\langle \sigma_{1,0}(x) \rangle =   \underset{\hbox{$(a)$}}{\vcenter{\hbox{\includegraphics[scale=0.8]{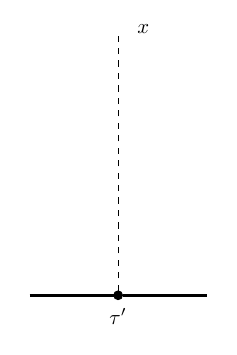}}} }\ \   + \ \  \underset{\hbox{$(b)$}}{\vcenter{\hbox{\includegraphics[scale=0.8]{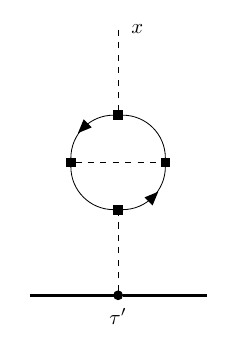}}}} \ \ + \ \  \underset{\hbox{$(b')$}}{\vcenter{\hbox{\includegraphics[scale=0.8]{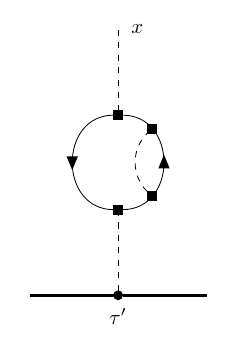}}}}  \ \   + \ \  \underset{\hbox{$(c)$}}{\vcenter{\hbox{\includegraphics[scale=0.8]{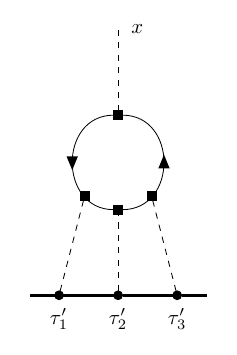}}}}.
\end{equation} 
The first diagram is the tree level contribution
\begin{equation}
(a) =  -h_0 \int d \tau' \langle\sigma_{1,0} (x) \sigma_{1,0}(\tau', \mathbf{0}) \rangle =  - \frac{h_0 \mathcal{N}_{\sigma} \pi }{|\mathbf{x}|}.  
\end{equation}
From the relation (\ref{LNORTHO}), we see that all the scalar legs in $(c)$ correspond to the auxiliary field $\sigma_{1}$. Thus, the contribution of this diagram can be obtained from the result derived in \cite{giombi2025linedefectsfermioniccfts} for the GNY model
\begin{equation}
(c) = \frac{ \tilde{g}_0^4 h_0^3 C(\delta) M^{-4\delta}}{|\mathbf{x}|^{1 + 4 \delta}} \approx \frac{h_0^3 \tilde{g}_0^4 }{|\mathbf{x}|} \left( \frac{c}{ N \delta} + \textrm{finite terms as $\delta \rightarrow 0$}  \right).
\end{equation}
\begin{equation} \label{LargeNc}
c =  -\frac{\mathcal{N}_{\sigma}^4 \pi^{4}   \Gamma \left(\frac{d - 2 }{2} \right)^2 \left[ (d - 3) \left(\psi \left(\frac{d}{2}-1\right)-\psi (d-3)\right) - 1  \right] }{12  (d - 3) \Gamma(d - 2) \sin \left(\frac{d \pi}{2} \right) } .
\end{equation}
For the diagrams in $(b)$ and $(b')$, the two legs outside the fermion loop correspond to $\sigma_{1}$, but the internal line can be any of the $N_s$ auxiliary fields $\sigma_{a}$. We have
\begin{equation}
    (b)\sim {\rm tr}_{flavor}{\rm tr}_{spinor}(\Sigma_1\gamma^\mu\Sigma_a\gamma^\nu\Sigma_1\gamma^\rho\Sigma_a\gamma^\lambda)
\end{equation}
\begin{equation}
    (b')\sim {\rm tr}_{flavor}{\rm tr}_{spinor}(\Sigma_1\gamma^\mu\Sigma_1\gamma^\nu\Sigma_a\gamma^\rho\Sigma_a\gamma^\lambda)
\end{equation}
From our assumptions about the nature of the $\Sigma_a$, we have that 
\begin{equation}    {\rm tr}_{flavor}{\rm tr}_{spinor}(\Sigma_1\gamma^\mu\Sigma_a\gamma^\nu\Sigma_1\gamma^\rho\Sigma_a\gamma^\lambda)\sim N(2-N_s)
\end{equation}
\begin{equation}
{\rm tr}_{flavor}{\rm tr}_{spinor}(\Sigma_1\gamma^\mu\Sigma_1\gamma^\nu\Sigma_a\gamma^\rho\Sigma_a\gamma^\lambda)\sim NN_s
\end{equation}
Let us parametrize the divergence part of the $(b)$ and $(b')$ diagrams as 
\begin{equation}
(b)  = \frac{h_0 \tilde{g}^4}{N \delta |\mathbf{x}|}b_1+\ldots .
\end{equation}
\begin{equation}
(b')  = \frac{h_0 \tilde{g}^4}{N  \delta |\mathbf{x}|}b_2+\ldots.
\end{equation}
One can see that these diagrams are the same that appear in the calculation of the $1/N$ correction to the scaling dimension of the bulk operators $\sigma_a$, which was computed in the GNY ($N_s=1$) and the NJLY ($N_s=2$) models in \cite{fei2017yukawacftsemergentsupersymmetry,1994PhRvD..50.2840G,Gracey_1993}. In terms of the coefficients $b_1$ and $b_2$, the scaling dimension reads 
\begin{equation}\label{eq:dsig}
	\Delta_\sigma=1-\frac{(b_1+b_2)}{N\pi\mathcal{N}_\sigma}+O(\frac{1}{N^2})
\end{equation}
Matching with the results in the literature \cite{fei2017yukawacftsemergentsupersymmetry,1994PhRvD..50.2840G,Gracey_1993}, we obtain explicitly 
\begin{equation}
    b_1=\pi\mathcal{N}_\sigma(2-N_s)(\alpha_1-\alpha_2)
\end{equation}
\begin{equation}
    b_2=\pi\mathcal{N}_\sigma N_s\alpha_2\,,
\end{equation}
where
\begin{equation}
    \alpha_1=\frac{2^{d+1}(d-1)\sin(\frac{\pi d}{2})\Gamma(\frac{d-1}{2})}{d(d-2)\pi^{\frac{3}{2}}\Gamma(\frac{d}{2}-1)}
\end{equation}
\begin{equation}
\alpha_2=\frac{2\Gamma(d-1)}{\Gamma(\frac{d}{2}-1)\Gamma(1-\frac{d}{2})\Gamma(1+\frac{d}{2})\Gamma(\frac{d}{2})}
\end{equation}

To cancel the $\frac{1}{\delta}$ divergence in (\ref{eq:sig10exp}), we must have
\begin{equation}
h_0 = M^{-\delta/2}h\left(1+\frac{ \tilde{g}^4 (b_1+b_2)}{2 N \delta \pi \mathcal{N}_{\sigma}} + \frac{h^2 \tilde{g}^4 c}{ N \delta \pi \mathcal{N}_{\sigma}}  + \dots\right)
\end{equation}
Therefore,  we obtain the large $N$ beta function
\begin{equation}
    \beta_{h} = -\frac{h (b_1+b_2)}{ N  \pi \mathcal{N}_{\sigma}} - \frac{3h^3 c}{ N  \pi \mathcal{N}_{\sigma}}
\end{equation}
One can see that the coefficient of $h$ in this beta function matches the anomalous dimension of $\sigma$ given in (\ref{eq:dsig}).
From the beta function, we obtain the non-trivial fixed point at large $N$ to be
\begin{equation}
    h_*^2=\frac{b_1+b_2}{3c}+O(\frac{1}{N})
\end{equation}

\begin{figure}[h!]
    \centering
    \includegraphics[width=0.5\linewidth]{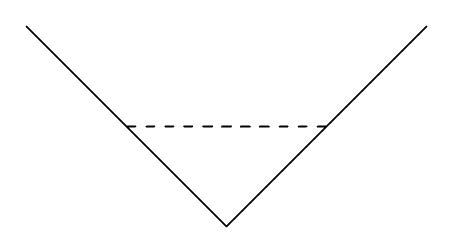}
    \caption{Diagram contributing to the cusp anomalous dimension to leading order at large $N$.}
    \label{LNCD}
\end{figure}

We can now move on to the calculation of the cusp anomalous dimension. We assume the two legs of the cusp are at an angle $\theta$ and that the unit vectors in $O(N_s)$ space along which the defect coupling vectors are oriented are $\hat{m}_1$ and $\hat{m}_2$. Using figure \ref{LNCD}, we obtain
\begin{equation}
    \Gamma_{\hat{m}_1\hat{m}_2}(\theta)=-h_*^2\mathcal{N}_\sigma \Big(\hat{m}_1\cdot\hat{m}_2\frac{\pi-\theta}{\sin\theta}-1\Big)
\end{equation}
Using the expressions of $\mathcal{N}_\sigma$ and $h_*^2$ given above, and setting $d=4-\epsilon$, we get
\begin{equation}
\label{eq:cusp-largeN}
    \Gamma_{\hat{m}_1\hat{m}_2}(\theta)=-\frac{1}{8\pi^2}\Big[4-N_s + \epsilon\Big(\frac{(4-N_s)\pi^2+3N_s-24}{12}\Big)\Big]\Big(\hat{m}_1\cdot\hat{m}_2\frac{\pi-\theta}{\sin\theta}-1\Big)+\mathcal{O}\Big(\frac{1}{N}\Big)
\end{equation}
To compare to the result obtained in the epsilon expansion calculations, note that, as shown in appendix \ref{app:appB}, the fixed point values of the various coupling constants at large $N$ take the form (see appendix \ref{app:appB})
\begin{equation}
    h_*^2=\frac{4-N_s}{2}+\epsilon\frac{(4-N_s)\pi^2+(24-9N_s)}{24} +\mathcal{O}\Big(\frac{1}{N}\Big)
\end{equation}
\begin{equation}
    g_{1*}^2=\frac{16\pi^2}{N}\epsilon + \mathcal{O}\Big(\frac{1}{N^2}\Big)
\end{equation}
\begin{equation}
    g_{2*}=\frac{192\pi^2}{N}+\mathcal{O}\Big(\frac{1}{N^2}\Big)
\end{equation}
 Plugging these values into (\ref{gengammaexp}) and keeping the leading large $N$ contributions, we find precise agreement with (\ref{eq:cusp-largeN}). 

\section{Results and Discussion}
\label{sec:results}
In this section, we discuss the calculation of several observables related to the cusp anomalous dimension computed above and discuss their extrapolation to $d=3$ ($\epsilon=1$). Within the models defined by the general theory (\ref{Sdef}), we focus on especially interesting examples with low $N$. In particular, we will consider the GNY model with $N=1$, and the NJLY model with $N=2$, which have been observed to have emergent supersymmetry at the IR fixed point \cite{Lee:2006if, Grover:2013rc, fei2017yukawacftsemergentsupersymmetry}. We will additionally look at the case with $N=2,N_s=1$, which corresponds to one Dirac fermion in $d=3$ coupled to a single real scalar. We tabulate some results in these cases at $d=3$ by setting $\epsilon=1$. 

\begin{figure}[h!]
    \centering
        \begin{subfigure}[b]{0.45\textwidth}
         \centering
         \includegraphics[width=\textwidth]{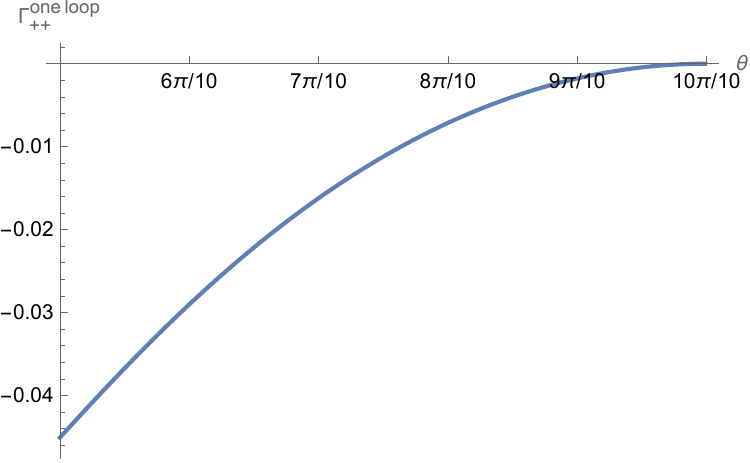}
         \caption{$\Gamma_{++}^{\text{one loop}}$}
         \label{gpp}
    \end{subfigure}
    \begin{subfigure}[b]{0.45\textwidth}
         \centering
         \includegraphics[width=\textwidth]{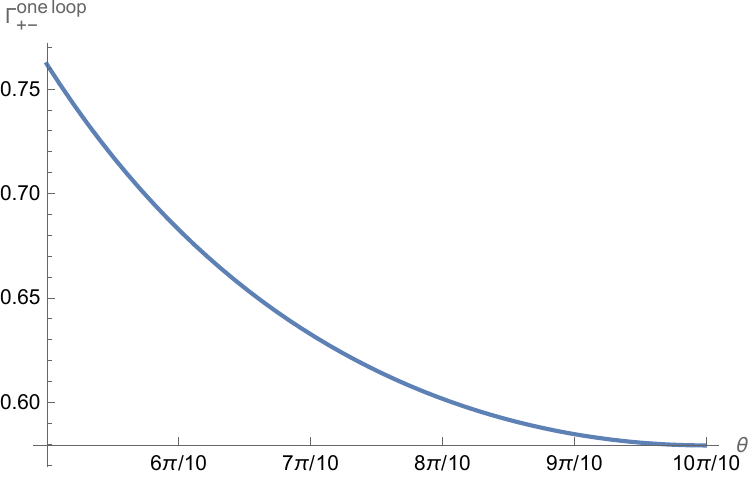}
         \caption{$\Gamma_{+-}^{\text{one loop}}$}
         \label{gpm}
    \end{subfigure}
    \caption{The term of $\mathcal{O}(\epsilon)$ of the cusp anomalous dimension in the two types of cusped defect in the GNY model.}
    \label{fig:cuspsplot1}
\end{figure}

\begin{figure}[h!]
	\centering
	\begin{subfigure}[b]{0.45\textwidth}
		\centering
		\includegraphics[width=\textwidth]{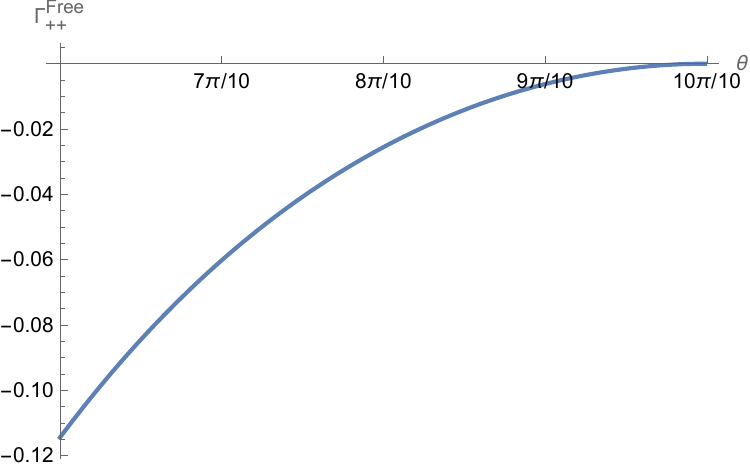}
		\caption{$\Gamma_{++}^{\text{free}}$}
		\label{gfpp}
	\end{subfigure}
	\begin{subfigure}[b]{0.45\textwidth}
		\centering
		\includegraphics[width=\textwidth]{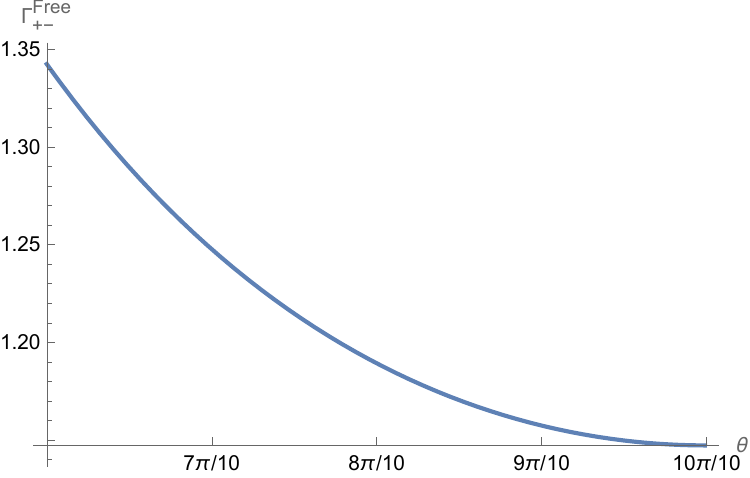}
		\caption{$\Gamma_{+-}^{\text{free}}$}
		\label{gfpm}
	\end{subfigure}
	\caption{The term of $\mathcal{O}(\epsilon^0)$ of the cusp anomalous dimension in the two types of cusped defect in the GNY model.}
	\label{fig:cuspsplotfree}
\end{figure}

\begin{figure}[h!]
    \centering
    \includegraphics[width=0.8\linewidth]{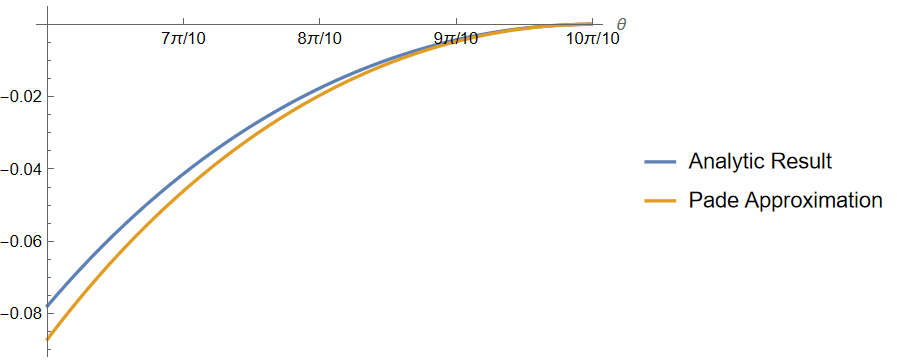}
    \caption{Pad\'e Approximation of $\Gamma_{++}(\theta)$ in the $d=3$ GNY model.}
    \label{fig:GNY_Pade}
\end{figure}

Focusing on the GNY model in the two cases $\hat{m}_1\cdot\hat{m}_2=\pm 1$ (equal or opposite couplings on the two arms of the cusp), we can visualize the angle dependence of the order $\epsilon$ term in the cusp anomalous dimension in figure \ref{fig:cuspsplot1} (the models with $N_s=2,3$ can be similarly considered). The $\mathcal{O}(\epsilon^0)$ term in the cusp anomalous dimension is visualized in figure \ref{fig:cuspsplotfree} for the two cases. In figure \ref{fig:GNY_Pade}, we compare the naive extrapolation of the epsilon expansion to $d=3$ with its Pad\'e approximation, focusing on the equal couplings case $\hat{m}_1\cdot\hat{m}_2= 1$. 

From the cusp anomalous dimension, we can extract various related quantities. In particular, we tabulate the numerical estimates for the dimensions of the defect changing and creation operators, as well as the Casimir energy in the various models mentioned earlier, in table \ref{tab:data1}. In the same table, we also include some numerical estimates for the dimension of the defect operator $\hat{\phi}$ using (\ref{dphihat}) and Pad\'e approximations. Note that if we set $N=0,N_s=1$, we obtain the Ising model, for which we obtain, using (\ref{dphihat}), the estimate $\Delta_{\hat{\phi}}^{(Ising)}=1.397$ in $d=3$ (this is not so close to the fuzzy sphere value of $\approx1.6$ \cite{Hu_2024,Zhou_2024}, it however agrees with the results obtained in \cite{Cuomo_2022} using a Pad\'e$_{(1,1)}$ approximation. The agreement can be improved \cite{Cuomo_2022} using a ``two-sided" Pade\'e approximant which incorporates a constraint given by the exact solution of the Ising model in $d=2$.). The computation of the quantities shown in table \ref{tab:data1}, as well as that of the $C_D$ and $C_t$ (for $N_s>1$) coefficients, is discussed in more detail in the following subsections.
\begin{table}[!ht]
    \centering
    {\renewcommand{\arraystretch}{1}%
    \begin{tabular}{|c|c|c|c|}
        \hline
         & GNY ($N=1$) & GNY ($N=2$) & NJLY ($N=2$) \\
         \hline
         \hline
         & & &\\
         $\Delta_{\hat{\phi}}$ (2-loop Pad\'e$_{(1,1)}$) & 1.3673 & 1.3386 & 1.3076\\
         \hline
         \hline
         & & &\\
         $\Delta_{+-}$  &\shortstack{0.8832 (One Loop) \\ 0.8836 (Pad\'e)} & \shortstack{0.7262 (One Loop)\\0.5821 (Pad\'e)} & \shortstack{0.4447 (One Loop)\\0.5819 (Pad\'e)}\\
         \hline 
         \hline
         & & &\\
         $\Delta_{\vec{h0}}$  & \shortstack{0.1851 (One Loop)\\0.1891 (Pad\'e)} & \shortstack{0.1553 (One Loop)\\0.1224 (Pad\'e)} &\shortstack{0.0926 (One Loop)\\0.1306 (Pad\'e)}\\
         \hline
         \hline
         & & &\\
         $\mathcal{E}_{++}$ ($\hat{m}_1\cdot\hat{m}_2=1$) & \shortstack{0.0782 (One Loop)\\-0.6616 (Pad\'e)} & \shortstack{-0.0639 (One Loop)\\-0.4100 (Pad\'e)} & \shortstack{-0.1508 (One Loop)\\-0.5249 (Pad\'e)} \\
         \hline
         \hline
         & & &\\
         $\mathcal{E}_{+-}$ ($\hat{m}_1\cdot\hat{m}_2=-1$) &\shortstack{1.4364 (One Loop)\\1.4408 (Pad\'e)} &\shortstack{1.1768 (One Loop)\\0.9536 (Pad\'e)} & \shortstack{0.7241 (One Loop)\\0.9372 (Pad\'e)}\\
         \hline
    \end{tabular}}
    \caption{Some results in $d=3$ for the GNY ($N_s=1$) and NJLY ($N_s=2$) models. Note that we have set $\hat{m}_1\cdot\hat{m}_2=-1$ in the calculation of the Defect Changing Operator.}
    \label{tab:data1}
\end{table}


\subsection{Dimension of Defect Changing and Creation (Annihilation) Operators}
Let us assume the defect coupling on both sides of the cusp to be of the form (\ref{defcoupspec}).
Using (\ref{eq:cadtodcodim}), we get the dimension of the defect changing operator to be 
\begin{equation}\label{DCOdim}
    \begin{split}
        \Delta_{\hat{m}_1,\hat{m}_2} =& -\frac{h_*^2}{4\pi^2}\left(\hat{m}_1\cdot\hat{m}_2-1\right) \\& -\frac{h_*^2}{4\pi^2}\left(\hat{m}_1\cdot\hat{m}_2-1\right)\left(-\frac{g_{1*}^2N}{16\pi^2}-\frac{g_{2*}h_*^2}{96\pi^2}\log(64 e^{-1})\right)\\&
        -\left(\epsilon\frac{h_*^2}{4\pi^2}-\frac{g_{1*}^2Nh_*^2}{64\pi^2}\right)((\hat{m}_1\cdot\hat{m}_2)-1)\\&
        +\frac{g_{2*}h_*^4}{768\pi^8}((\hat{m}_1\cdot\hat{m}_2)-1)(8\pi^4-4\pi^4\log8-24\pi^2\zeta(3))\\&
        +\frac{g_{2*}h_*^4}{1024\pi^8}(\left(\frac{1+2(\hat{m}_1\cdot\hat{m}_2)^2}{3}\right)-1)(24\pi^2\zeta(3))
    \end{split}
\end{equation}
More compactly,
\begin{equation}\label{DCOdim2}
\begin{split}
    \Delta_{\hat{m}_1,\hat{m}_2}=& (1-\hat{m}_1\cdot\hat{m}_2)\Big[\frac{h_*^2}{4\pi^2}(1+\epsilon)-\frac{g_{1*}^2Nh_*^2}{32\pi^4}\\&-\frac{g_{2*}h_*^4}{128\pi^6}(\pi^2-4\zeta(3))+\frac{g_{2*}h_*^4\zeta(3)}{64\pi^6}(1+\hat{m}_1\cdot\hat{m}_2) \Big]
\end{split}
\end{equation}
Note that, as expected, we have an overall factor of $(1-\hat{m}_1\cdot\hat{m}_2)$, since for $\hat{m}_1\cdot \hat{m}_2 = 1$ there is no defect changing operator. We also checked that the expression reduces to the one obtained in \cite{cuomo2024impuritiescuspgeneraltheory} in the limit as $N\rightarrow0$.

We may write
\begin{equation}
\Delta_{\hat{m}_1\hat{m}_2}=\Delta_{\hat{m}_1\hat{m}_2}^{(0)}+\epsilon\Delta_{\hat{m}_1\hat{m}_2}^{(1)}+\mathcal{O}(\epsilon^2)
\end{equation}
Substituting the fixed point values from appendix \ref{app:appB}, we obtain
\begin{equation}
    \begin{split}
\Delta_{\hat{m}_1\hat{m}_2}^{(0)}=(1-\hat{m}_1\cdot\hat{m}_2)\frac{(4-N_s)(N_s+8)}{\pi^2\Big( 8 - N - 2 N_s + \sqrt{ N^2 + 4 (-4 + N_s)^2 + 4 N (28 + 5 N_s) } \Big)}
    \end{split}
\end{equation}
and
\begin{equation}
    \begin{split}        \Delta_{\hat{m}_1\hat{m}_2}^{(1)}=&(1-\hat{m}_1\cdot\hat{m}_2)\Bigg[\frac{h_{*,1}^2}{4\pi^2}+\frac{h_{*,0}^2}{4\pi^4}\Bigg(\frac{\pi^2(N_2-N-4)+6(4-N_s)(3+\hat{m}_1\cdot\hat{m}_2)\zeta(3)}{(8+N-2N_s)}\Bigg)\Bigg]
    \end{split}
\end{equation}

Now, let us take $\vec{h}_2=0$ and $\vec{h}_1=\vec{h}=h_*\hat{m}$, which corresponds to the case of creation (or annihilation) of the defect. Thus, using \ref{gengammaexpmixed}, we get
\begin{equation}
    \Delta_{0,\vec{h}}=\Gamma_{0\vec{h}}(\pi)
\end{equation}
Note that the dimensions of the defect creation and annihilation operators are of course the same 
\begin{equation}
    \Delta_{0\vec{h}}=\Delta_{\vec{h}0}
\end{equation}
We obtain 
\begin{equation}
	\Delta_{0,\vec{h}}=\frac{h_{*}^2}{8\pi^2}(1+\epsilon)-\frac{g_{1*}^2Nh_{*,0}^2}{64\pi^4}        -\frac{g_{2*}h_{*,0}^4}{256\pi^6}\Big(\pi^2-\zeta(3)\Big)+\mathcal{O}(\epsilon^2)
\end{equation}
As before, let us write the epsilon expansion as
\begin{equation}
\Delta_{0,\vec{h}}=\Delta_{0,\vec{h}}^{(0)}+\epsilon\Delta_{0,\vec{h}}^{(1)}+\mathcal{O}(\epsilon^2)
\end{equation}
Substituting the fixed point values of the bulk and defect couplings, we obtain
\begin{equation}
    \Delta_{0,\vec{h}}^{(0)}=\frac{h_{*,0}^2}{8\pi^2}=\frac{(4-N_s)(N_s+8)}{2\pi^2\Big( 8 - N - 2 N_s + \sqrt{ N^2 + 4 (-4 + N_s)^2 + 4 N (28 + 5 N_s) } \Big)}
\end{equation}
\begin{equation}
    \begin{split}        \Delta_{0,\vec{h}}^{(1)}=\frac{h_{*,1}^2}{8\pi^2}+\frac{h_{*,0}^2}{8\pi^4}\Bigg(\frac{\pi^2(N_s-N-4)+3(4-N_s)\zeta(3)}{(8+N-2N_s)}\Bigg)
    \end{split}
\end{equation}

\subsection{Casimir Energy}
\label{sec:casimir}
As discussed in the introduction, we can also compute the Casimir energy related to fusion of the defects by looking at the expansion of the cusp anomalous dimension \ref{gengammaexp} around $\theta=0$. The details of the expansion around $\theta=0$ are given in appendix \ref{app:appA2}. We use the following ansatz for this expansion, following \cite{cuomo2024impuritiescuspgeneraltheory} 
\begin{equation}\label{t0exp}
\Gamma_{\hat{m}_1\hat{m}_2}(\theta\rightarrow0)=\frac{\mathcal{E}_{\hat{m}_1\hat{m}_2}}{\theta}+\delta\log\theta+\alpha+\dots
\end{equation}
In this expression, $\mathcal{E}_{\hat{m}_1\hat{m}_2}$ denotes the Casimir energy (in the notation of eq. (\ref{casimirdefcad}), $\mathcal{E}_{\hat{m}_1\hat{m}_2} = C_{\vec{h}_1 \vec{h}_2 \vec{h}_{\rm fus}}$). According to (\ref{casimirdefcad}), the constant term $\alpha$ is expected to be related to the defect creation dimension $\Delta_{\vec{h}_{\rm fus}0}$ of the fused line, while the $\log\theta$ term arises in the perturbative epsilon expansion because the dimension of the irrelevant operator in (\ref{casimirdefcad}) is $\Delta_{\rm irr}=1+O(\epsilon)$. At the classical level, the the fused defect coupling is $\vec{h}_{\rm fus}= \vec{h}_1+\vec{h}_2$, which is not at the fixed point (since $|\vec{h}_1^*+\vec{h}_2^*| \neq h^*$ generically). The RG flow is expected to drive the defect coupling to the fixed point value $\frac{\hat{m}_1+\hat{m}_2}{\sqrt{2+2\hat{m}_1\cdot \hat{m}_2}}h_*$ \cite{diatlyk2024defectfusioncasimirenergy, cuomo2024impuritiescuspgeneraltheory}. 

At small $\theta$, starting from the expression in (\ref{gengammaexp}), we can extract the $\frac{1}{\theta}$ coefficient to be
\begin{equation}\label{cadsmth}
\begin{split}
    \Gamma_{h_*\hat{m}_1,h_*\hat{m}_2}\sim\frac{1}{\theta}\Bigg[& -\frac{h_*^2(\hat{m}_1\cdot\hat{m}_2)}{4\pi}-\frac{h_*^2(\hat{m}_1\cdot\hat{m}_2)}{4\pi}\left(-\frac{g_{1*}^2N}{16\pi^2}-\frac{g_{2*}h_*^2}{96\pi^2}\log(64 e^{-1})\right) \\& -\left(\epsilon\frac{h_*^2}{8\pi^2}-\frac{g_{1*}^2Nh_*^2}{128\pi^4}\right)2\pi(\hat{m}_1\cdot\hat{m}_2)\log2+\frac{g_{2*}h_*^4}{96\pi^3}(1-\log2)(\hat{m}_1\cdot\hat{m}_2)\\&+\frac{g_{2*}h_*^4}{1024\pi}\left(\frac{1+2(\hat{m}_1\cdot\hat{m}_2)^2}{3}\right)\Bigg]
\end{split}
\end{equation}
Using \ref{casimirdefcad}, we obtain the Casimir energy to be
\begin{equation}\label{casimir}
    \begin{split}
        \mathcal{E}_{h_*\hat{m}_1,h_*\hat{m}_2} =\Bigg[& -\frac{h_*^2(\hat{m}_1\cdot\hat{m}_2)}{4\pi}-\frac{h_*^2(\hat{m}_1\cdot\hat{m}_2)}{4\pi}\left(-\frac{g_{1*}^2N}{16\pi^2}-\frac{g_{2*}h_*^2}{96\pi^2}\log(64 e^{-1})\right) \\& -\left(\epsilon\frac{h_*^2}{8\pi^2}-\frac{g_{1*}^2Nh_*^2}{128\pi^4}\right)2\pi(\hat{m}_1\cdot\hat{m}_2)\log2+\frac{g_{2*}h_*^4}{96\pi^3}(1-\log2)(\hat{m}_1\cdot\hat{m}_2)\\&+\frac{g_{2*}h_*^4}{1024\pi}\left(\frac{1+2(\hat{m}_1\cdot\hat{m}_2)^2}{3}\right)\Bigg]
    \end{split}
\end{equation}
Let us write
\begin{equation} \mathcal{E}_{h_*\hat{m}_1,h_*\hat{m}_2}=\mathcal{E}_{h_*\hat{m}_1,h_*\hat{m}_2}^{(0)}+\epsilon\mathcal{E}_{h_*\hat{m}_1,h_*\hat{m}_2}^{(1)}+\mathcal{O}(\epsilon^2)
\end{equation}
Then, substituting the fixed point values from appendix \ref{app:appB}, we get
\begin{equation}
\mathcal{E}_{h_*\hat{m}_1,h_*\hat{m}_2}^{(0)}=-\frac{(4-N_s)(N_s+8)}{\pi\Big( 8 - N - 2 N_s + \sqrt{ N^2 + 4 (-4 + N_s)^2 + 4 N (28 + 5 N_s) } \Big)}(\hat{m}_1\cdot\hat{m}_2)
\end{equation}
\begin{equation}
    \begin{split}        \mathcal{E}_{h_*\hat{m}_1,h_*\hat{m}_2}^{(1)}=&-\frac{h_{*,1}^2(\hat{m}_1\cdot\hat{m}_2)}{4\pi}\\&+\frac{h_{*,0}^2}{4\pi}\Bigg(\frac{(\hat{m}_1\cdot\hat{m}_2)(N+12-3N_s)}{(N+8-2N_s)}\\&+\frac{(4-N_s)}{(8+N-2N_s)}\Big[\frac{(1+2(\hat{m}_1\cdot\hat{m}_2)^2)\pi^2}{8}\Big]\Bigg)
    \end{split}
\end{equation}
This expression reduces to the correct form for the GNY and $O(N)$ models as computed previously in \cite{diatlyk2024defectfusioncasimirenergy}. In the above, we used (\ref{gengammaexp}) where the defect couplings are taken to be non-trivial on both lines. If we considered the case where $\vec{h}_1=h_*\hat{m}_1$ and $\vec{h}_2=0$, which can be obtained from (\ref{gengammaexpmixed}), we would find instead that the Casimir energy is zero as expected from the fact that the defect with $\vec{h}=0$ is trivial and acts as the identity element when fusing defects.    

For the subleading terms in (\ref{t0exp}), we obtain
\begin{equation}    \delta=\frac{g_{2*}h_{*}^4}{192\pi^4}(1+3(\hat{m}_1\cdot\hat{m}_2)+2(\hat{m}_1\cdot\hat{m}_2)^2)
\end{equation}
Substituting the fixed point values, we obtain
\begin{equation}    \delta=\frac{\epsilon2(4-N_s)^2(8+N_s)(1+3(\hat{m}_1\cdot\hat{m}_2)+2(\hat{m}_1\cdot\hat{m}_2)^2)}{\pi^2(8+N-2N_S)\Big( 8 - N - 2 N_s + \sqrt{ N^2 + 4 (-4 + N_s)^2 + 4 N (28 + 5 N_s) } \Big)}
\end{equation}
Upon calculating $\alpha$, we observe that it is simply what we would obtain if we substitute $\vec{h}_2=0$ and $\vec{h}_1=h_*(\hat{m}_1+\hat{m}_2)$ in (\ref{gengammaexpmixed}) to $\mathcal{O}(\epsilon^0)$, giving the dimension of the defect creation operator of the fused defect to leading order. This confirms the expected structure (\ref{casimirdefcad}) to $\mathcal{O}(\epsilon^0)$. Similarly to what was found in \cite{cuomo2024impuritiescuspgeneraltheory} in the analogous case of the pinning field defect of the $O(N)$ model, we observe that $\delta=0$ if $\hat{m}_1\cdot\hat{m}_2=-\frac{1}{2}$, which means that the $\log$ term in (\ref{t0exp}) vanishes in this case. This is expected because in such case the classical value of the fused defect coupling $\vec{h}_{\rm fus}= \vec{h}_1+\vec{h}_2$ is already at its expected fixed point value. For this special case of $\hat{m}_1\cdot\hat{m}_2=-\frac{1}{2}$, one can also verify that the constant term $\alpha$ in fact agrees with the defect creating operator dimension to $O(\epsilon)$.  


\subsection{Two point functions of the displacement and tilt operators}\label{disptiltsec}
\begin{figure}[h!]
    \centering
    \includegraphics[width=0.5\linewidth]{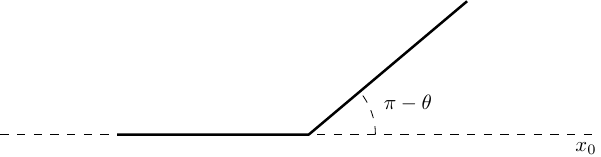}
    \caption{Setup for the calculation of  the two point function of the displacement operator.}
    \label{fig:dispcuspset}
\end{figure}

As shown in \cite{Correa:2012at, Cavagli__2023, cuomo2024impuritiescuspgeneraltheory}, one has the following relation between the quadratic term in the near-straight expansion of the cusp anomalous dimension and the integrated two-point function of the displacement operator on the straight line
\begin{equation}
    \frac{d^2}{d\theta^2}\Gamma_{\hat{m}\hat{m}}(\theta)\Big|_{\theta=\pi} \log\Big(\frac{L}{a}\Big)=-\int_0^\infty d\tau_1d\tau_2\text{ }\tau_1\tau_2\langle D_n(\tau_1)D_n(\tau_2)\rangle_c
\end{equation}
Here $D_n(\tau)$ is a displacement operator \cite{Cuomo_2022_g} in the direction $n$ orthogonal to the defect contour inserted at a point that is a distance $\tau$ from the cusp, and the two insertions are on the $\tau>0$ half-line. One can derive this relation by looking at the cusped defect as shown in figure \ref{fig:dispcuspset}, viewing it as a deformation of the infinite straight line. Taking equal couplings on the two arms of the cusp, we can write the logarithm of the expectation value of the defect as
\begin{equation}
	\log \frac{Z_{\hat{m}\hat{m}}(\theta)}{Z_{CFT}}=\log\Bigg[\int [D\vec{\phi} D\Psi]e^{-S_0}e^{-\delta S}\Bigg]
\label{Zdeform}
\end{equation} 
where $S_0$ is the action with the undeformed defect ($\theta=\pi$), and the deformation reads, to the order we are concerned with
\begin{equation}
\label{deltaS-D}
	\delta S= (\pi-\theta)\int_0^\infty d\tau \tau D(\tau) + (\pi-\theta)^2\int_0^\infty d\tau\Big(\frac{b}{a^3}\tau^2\mathbb{1}(\tau)\Big)+\dots 
\end{equation}
Here, the identity operator is inserted to cancel out divergences coming from the region where the integrated points collide, and $a$ is a short-distance regulator (defined so that the integrated points satisfy $|x-y| > a$).  

Recall that, since the displacement operator is protected and has $\Delta(D_n)=2$, its two-point function on the straight line is given by  
\begin{equation}
    \langle D_n(\tau_1)D_n(\tau_2)\rangle=\frac{C_D}{|\tau_1-\tau_2|^4}
\end{equation}
where the normalization factor $C_D$ is an observable of the dCFT. Plugging this into (\ref{Zdeform}) and using the definition (\ref{flspcadded}), we find the relation 
\begin{equation}
    \frac{d^2}{d\theta^2}\Gamma_{\hat{m}\hat{m}}(\theta)\Big|_{\theta=\pi} \log\Big(\frac{L}{a}\Big)=-\int_{0}^\infty d\tau_1d\tau_2 \text{ }\tau_1\tau_2\frac{C_D}{(\tau_1-\tau_2)^4}-2b_1\frac{L^3}{3a^3}
\end{equation}
The constant $b$ can be fixed to cancel out the divergence of the form $\frac{L^3}{a^3}$ that comes from the first integral. This gives $b=-\frac{C_D}{3}$. Comparing the coefficient of the logarithmic divergence on both sides of the above equation, one finds the relation
\begin{equation}\label{CD}
    \frac{d^2}{d\theta^2}\Gamma_{\hat{m}\hat{m}}(\theta)\Big|_{\theta=\pi} =-\frac{C_D}{6}
\end{equation}
Expanding the various functions appearing in the anomalous dimension near $\theta=\pi$ (see appendix \ref{app:appA2} for the details), we get
\begin{equation}\label{cadthpi}
    \begin{split}
        \Gamma''(\pi)&=-\frac{h_*^2}{12\pi^2}- \frac{h_*^2}{12\pi^2}\left(-\frac{g_{1*}^2N}{16\pi^2}-\frac{g_{2*}h_*^2}{96\pi^2}\log(64 e^{-1})\right)\\&-\frac{2}{9}\left(\epsilon\frac{h_*^2}{8\pi^2}-\frac{g_{1*}^2Nh_*^2}{128\pi^2}\right)+\frac{g_{2*}h_*^2}{1152\pi^8}(4\pi^4-2\pi^2-2\pi^4\log8-12\pi^2\zeta(3))\\&+\frac{g_{2*}h_*^4}{512\pi^8}\sigma_2
    \end{split}
\end{equation}
Upon using \footnote{The derivation of this is given in appendix \ref{app:appA2}. This analytic result agrees with the numerical value for $\sigma_2$ given in \cite{cuomo2024impuritiescuspgeneraltheory}.}
\begin{equation}\label{sig2analytic}
    \sigma_2=\frac{8\pi^2}{9}(1+6\zeta(3))
\end{equation}
we get
\begin{equation}\label{CDFullexp}
    \begin{split}
        C_D  &=\frac{h_*^2}{2\pi^2}+ \frac{h_*^2}{2\pi^2}\left(-\frac{g_{1*}^2N}{16\pi^2}-\frac{g_{2*}h_*^2}{96\pi^2}\log(64 e^{-1})\right)\\&+\frac{4}{3}\left(\epsilon\frac{h_*^2}{8\pi^2}-\frac{g_{1*}^2Nh_*^2}{128\pi^2}\right)-\frac{g_{2*}h_*^4}{192\pi^8}(4\pi^4-2\pi^2-2\pi^4\log8-12\pi^2\zeta(3))\\&+\frac{g_{2*}h_*^4}{96\pi^6}(1+6\zeta(3))    
    \end{split}
\end{equation}
Let us write 
\begin{equation}\label{NormalD}
    C_D=h_*^2\mathcal{N}_D^2
\end{equation}
Using the fixed point values of the couplings, we end up with 
\begin{equation}\label{NormalDfullexp}
\mathcal{N}_D^2=\frac{1}{2\pi^2}\Big(1+\epsilon\Big[\frac{4-N_s}{3(8+N-2N_s)}\Big]\Big)
\end{equation}
This precisely matches the expression obtained in \cite{barrat2025linedefectcorrelatorsfermionic}. If we set $N=0$, we obtain the same expression as calculated in \cite{Gimenez_Grau_2022} for the pinning field defect in the $O(N_s)$ model.

Using the one-loop epsilon expansion results and Pad\'e resummants, one may obtain the following numerical estimates for $C_D$ in some interesting special models in $d=3$
\begin{table}[!ht]
    \centering
    {\renewcommand{\arraystretch}{1}%
    \begin{tabular}{|c|c|c|c|}
        \hline
         & $N=1,N_s=1$ & $N=2,N_s=1$ & $N=2,N_s=2$ \\
         \hline
         & & &\\
         $C_D$ (One loop $\epsilon$ expansion) &0.9100  & 0.7778 &0.6387\\
         \hline
         & & &\\
         $C_D$ (Pad\'e) &0.9122  & 0.7792 & 0.6396\\
         \hline
    \end{tabular}}
    \caption{Numerical values for $C_D$ in $d=3$, for some special models.}
    \label{tab:data2}
\end{table}
In models with $N_s>1$, we have an additional set of protected operators with $\Delta=1$ which arise from the breaking of $O(N_s)$ to $O(N_s-1)$ due to the defect. They are known as tilt operators, and their two-point function is given by 
\begin{equation}
\langle t^a(\tau_1)t^b(\tau_2)\rangle = \frac{C_t}{|\tau_1-\tau_2|^2}\delta^{ab},\qquad a,b=1,\ldots, N_s-1
\end{equation}
For convenience, let us denote 
\begin{equation}
    \hat{m}_1\cdot\hat{m}_2=\cos\theta_f
\end{equation}
where $\theta_f$ is an angle between vectors in the internal $O(N_s)$ space.
\begin{figure}
    \centering
    \includegraphics[width=0.5\linewidth]{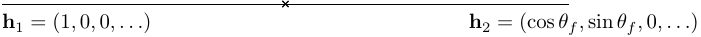}
    \caption{Setup for the calculation of the two point function of the tilt operator.} 
    \label{fig:tiltcuspst}
\end{figure}
We consider a setup as shown in figure \ref{fig:tiltcuspst} to calculate the two point function of the tilt operator from the cusp anomalous dimension. As discussed above for the displacement operator, we can view this setup as a  deformation of the straight line with uniform coupling along the whole line. Similarly to (\ref{deltaS-D}), the deformation action in this case can be written as 
\begin{equation}
	\delta S=\theta_f\int_0^\infty d\tau t(\tau)+\theta_f^2\int_0^\infty d\tau \Big[\frac{b}{a}\mathbb{1}(\tau)\Big]+\ldots 
\end{equation}
where $b$ again parametrizes a counterterm needed to cancel power divergences, and $t(\tau)$ denotes the tilt operator in a specific direction. 
Putting everything together, we have
\begin{equation}
\frac{\partial^2\Gamma(\theta,\theta_f)}{\partial\theta_f^2}\Bigg|_{\theta_f=0,\theta=\pi}\log\Big(\frac{L}{a}\Big)=-\int_0^\infty\int_0^\infty d\tau_1d\tau_2 \frac{C_t}{(\tau_1-\tau_2)^2} -b\frac{L}{a}
\end{equation}
By requiring the $\frac{L}{a}$ divergence get canceled, one gets $b=-C_t$ \cite{Cavagli__2023}. We thus obtain
\begin{equation}
\frac{\partial^2\Gamma(\theta,\theta_f)}{\partial\theta_f^2}\Bigg|_{\theta_f=0,\theta=\pi}=C_t
\end{equation}
Let us write for convenience 
\begin{equation}
C_t = h_*^2\mathcal{N}_t^2
\end{equation}
Substituting the explicit expression we obtained for $\Gamma(\theta,\theta_f)$, we get
\begin{equation}
\begin{split}
    h_*^2\mathcal{N}_t^2&=\frac{h_*^2}{4\pi^2}\\&
    +\frac{h_*^2}{4\pi^2}\left(-\frac{g_{1*}^2N}{16\pi^2}-\frac{g_{2*}h_*^2}{96\pi^2}\log(64 e^{-1})\right)\\&+\left(\epsilon\frac{h_*^2}{4\pi^2}-\frac{g_{1*}^2Nh_*^2}{64\pi^2}\right)\\&+\frac{g_{2*}h_*^4}{768\pi^8}(-8\pi^4+4\pi^4\log8+24\pi^2\zeta(3))\\&-\frac{g_{2*}h_*^4}{32\pi^6}\zeta(3)
\end{split}
\end{equation}
Using the fixed point values of the couplings, we get
\begin{equation}\label{tnn}    \mathcal{N}_t^2=\frac{1}{4\pi^2}\Big(1-\epsilon\Big[1+\frac{N_s-4}{8+N-2N_s}\Big]\Big)
\end{equation}
which agrees with the result found in \cite{barrat2025linedefectcorrelatorsfermionic}. 
Note that we need at least $N_s=2$ to make sense of $C_t$\footnote{In the $N_s=1$ case, we have no tilt operators, and so the result we obtain by direct substitution in (\ref{tnn}) is not meaningful.}.
For the $N=2,N_s=2$ model in $d=3$, we obtain $C_t$ to be $0.2405$ using the one loop $\epsilon$ expansion, and $0.2527$ using a Pad\'e approximation. 

Higher-order terms in the expansion of $\Gamma(\theta,\theta_f)$ near $\theta=\pi$, $\theta_f=0$ will encode in a similar way integrated higher-point functions of the displacement and tilt operators. In the case of the half-BPS Wilson line in ${\cal N}=4$ SYM, ref.~\cite{Cavagli__2023} derived non-trivial identities relating the cusp anomalous dimension to integrated four-point functions. A similar approach can be applied to the models studied in this paper. We leave the study of the resulting integrated correlator constraints to future work. 

\section*{Acknowledgments}
This work was supported in part by the US National Science Foundation Grant No. PHY-2209997. We thank G. Cuomo and O. Diatlyk for useful discussions.  
\appendix
\section{Detailed Calculations}\label{app:appA}

\subsection{Calculation of the fermion loop diagram}\label{app:appA1}
\begin{figure}[h!]
    \centering
    \includegraphics[width=0.5\linewidth]{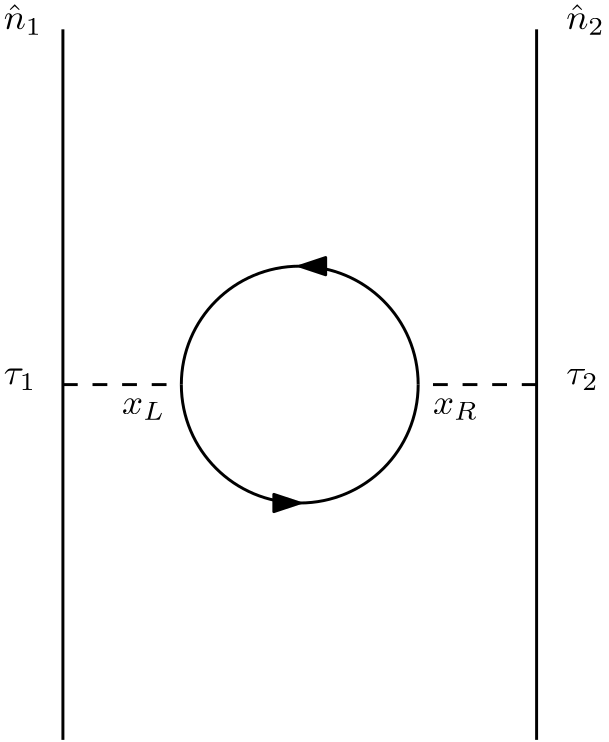}
    \caption{Fermion loop diagram.}
    \label{fig:A1}
\end{figure}
We use the same notation as in \cite{giombi2025linedefectsfermioniccfts} where $g_1$ and $g_2$ denote the Yukawa and quartic bulk couplings respectively. We use the same notations as used in \cite{cuomo2024impuritiescuspgeneraltheory} to denote the relevant integrals appearing in the calculation. To summarize all relevant notations, we define
\begin{equation}\label{eq:fdef}
    f_d(x)=\int d\tau \frac{\exp\big({-(\frac{d}{2}-1)|\tau|}\big)}{\big(1+\exp(-2|\tau|)-2x\exp(-|\tau|)\big)^{\frac{d}{2}-1}}
\end{equation}
\begin{equation}\label{eq:Nddef}
    \mathcal{N}_d=\frac{\Gamma\big(\frac{d}{2}-1\big)}{4\pi^\frac{d}{2}}
\end{equation}
\begin{equation}\label{eq:Fdef}
    F_{\epsilon}(x)=\int d\tau \frac{ \exp(-(1-\epsilon)|\tau|)}{(1-2x\exp(-|\tau|)+\exp(-2|\tau|))^{1-\epsilon}}
\end{equation}
\begin{equation}\label{eq:F'def}
    F'_a=\frac{dF_{\epsilon}}{d\epsilon} |_{ \epsilon=a}
\end{equation}
\begin{equation}\label{eq:F0val}
    F_0(\cos\theta)=\frac{\pi-\theta}{\sin\theta}
\end{equation}
\begin{equation}\label{eq:fFrel}
    f_{4-\epsilon}(x)=F_{\frac{\epsilon}{2}}(x)
\end{equation}

On the cylinder, we have the propagators to be \cite{Birrell_Davies_1982}
\begin{equation}
    G_{scalar}(\tau,\hat{n}_1;0,\hat{n}_2)=\frac{\Gamma(\frac{d}{2}-1)}{4\pi^{\frac{d}{2}}}\frac{\exp(-\frac{d-2}{2}|\tau|)}{(1-2\hat{n}_1\cdot\hat{n}_2 e^{-|\tau|}+e^{-2|\tau|})^{\frac{d-2}{2}}}
\end{equation}
\begin{equation}
G_{fermion}(i,\tau,\hat{n}_1;j,0,\hat{n}_2)=\delta_{ij}(d-2)\frac{\Gamma(\frac{d}{2}-1)}{4\pi^{\frac{d}{2}}}\frac{\exp(-\frac{d-1}{2}|\tau|)}{(1-2\hat{n}_1\cdot\hat{n}_2 e^{-|\tau|}+e^{-2|\tau|})^{\frac{d-1}{2}}}\gamma^\mu(\Hat{x_1-x_2})_{\mu}
\end{equation}
\\($(\Hat{x_1-x_2})_{\mu}$ is a unit vector joining $x_1$ and $x_2$). 

We can compute the diagram in figure \ref{fig:A1} (denoted as $GNY$) as 
\begin{equation}
\begin{split}
GNY&=-h_{1,0}h_{2,0}g_{1,0}^{2}\left(\frac{\Gamma(\frac{d}{2}-1)}{4\pi^{\frac{d}{2}}}\right)^4 (d-2)^2 N\int d\tau_1\int d\tau_2\int d\tau_L\int d\tau_R \int d^{d-1}\hat{n}_L\int d^{d-1}\hat{n}_R \\& \frac{\exp(-\frac{d-2}{2}|\tau_1-\tau_L|)}{(1-2\hat{n}_1\cdot\hat{n}_L e^{-|\tau_1-\tau_L|}+e^{-2|\tau_1-\tau_L|})^{\frac{d-2}{2}}}\frac{\exp(-\frac{d-2}{2}|\tau_2-\tau_R|)}{(1-2\hat{n}_R\cdot\hat{n}_2 e^{-|\tau_2-\tau_R|}+e^{-2|\tau_2-\tau_R|})^{\frac{d-2}{2}}}\\&\frac{\exp(-(d-1)|\tau_L-\tau_R|)}{(1-2\hat{n}_L\cdot\hat{n}_R e^{-|\tau_L-\tau_R|}+e^{-2|\tau_L-\tau_R|})^{d-1}}
\end{split}
\end{equation}
We can change variable among the $\tau$'s and we obtain
\begin{equation}\label{gnycyl}
\begin{split}
    GNY&=T [-h_{1,0}h_{2,0}g_{1,0}^{2}\left(\frac{\Gamma(\frac{d}{2}-1)}{4\pi^{\frac{d}{2}}}\right)^4 (d-2)^2  N\int d\tau_1\int d\tau_2\int d\tau \int d^{d-1}\hat{n}_L\int d^{d-1}\hat{n}_R \\& \frac{\exp(-\frac{d-2}{2}|\tau_1|)}{(1-2\hat{n}_1\cdot\hat{n}_L e^{-|\tau_1|}+e^{-2|\tau_1|})^{\frac{d-2}{2}}}\frac{\exp(-\frac{d-2}{2}|\tau_2|)}{(1-2\hat{n}_R\cdot\hat{n}_2 e^{-|\tau_2|}+e^{-2|\tau_2|})^{\frac{d-2}{2}}}\\&\frac{\exp(-(d-1)|\tau|)}{(1-2\hat{n}_L\cdot\hat{n}_R e^{-|\tau|}+e^{-2|\tau|})^{d-1}} ]
\end{split}
\end{equation}
Before going ahead with computing this complicated looking expression, it would be beneficial to understand what the diagram would look like if we do the calculation in flat space. In flat space, the value of $\langle s(x)s(0)\rangle$ with a single fermion loop i.e. the one loop correction to the scalar propagator, is
\begin{equation}\label{ss1l}
   \langle s(x)s(0)\rangle_{\text{1-Loop}} = \frac{g_1^2N}{64\pi^d}\frac{\Gamma(\frac{d}{2}-1)^2 \Gamma(d-3)}{\Gamma(d-2) (4-d)}\frac{1}{(x^2)^{d-3}} 
\end{equation}
If we naively convert this expression into one on the cylinder and use it to compute the diagram in figure \ref{fig:A1}, we expect the divergent part to match that obtained from the full calculation on the cylinder. Showing that the next term matches requires some effort.

Let us now introduce some further notation. Define the following
\begin{equation}\label{gdeltdef}
    G^\Delta_{\rm cyl}(t, \omega_1; 0, \omega_2) = \bigl( \cosh t - \omega_1 \cdot \omega_2 \bigr)^{-\Delta}
\end{equation}
This is the most general form of the propagators we get in the cylinder frame (dropping all the constants out front).
We have the following expansion \cite{giombi2025higherloopsadsapplications}
\begin{equation}
    G^\Delta_{\rm cyl}(t, \omega_1; 0, \omega_2) = \sum_{\ell m} \int_{\mathbb R} \frac{d\nu}{2\pi} \, K_\Delta(\nu,\ell) \, e^{i\nu t}\, Y_{\ell m}(\omega_1) Y^*_{\ell m}(\omega_2)
\end{equation}
Where
\begin{align}
    K_\Delta(\nu,\ell) &= \frac{  2^{2+\alpha} \, \pi^{\frac{d}{2}} }{\Gamma(\Delta)}  \int_0^\infty \frac{ds}{s} \, s^{\Delta-\alpha} \, K_{i \nu}(s) \, I_{\alpha+\ell}(s) \, \\
 &= \frac{2^\Delta \pi^{\frac{d}{2}} \Gamma(\frac{d}{2}-\Delta)}{\Gamma(\Delta)}
  \frac{\Gamma\bigl( \frac{1}{2}(\Delta+\ell-i \nu) \bigr) \, \Gamma\bigl( \frac{1}{2}(\Delta+\ell+i \nu) \bigr)}{\Gamma\bigl( \frac{1}{2}(\bar\Delta+\ell+i \nu) \bigr) \, \Gamma\bigl( \frac{1}{2}(\bar\Delta+\ell-i \nu) \bigr)}
\end{align}
Here, we integrate over the defect i.e. the parameter $t$, and thus, we can set $\nu=0$. Thus, we get
\begin{align}
     K_\Delta(0,\ell) &= \frac{  2^{2+\alpha} \, \pi^{\frac{d}{2}} }{\Gamma(\Delta)}  \int_0^\infty \frac{ds}{s} \, s^{\Delta-\alpha} \, K_{0}(s) \, I_{\alpha+\ell}(s) \, \\
 &= \frac{2^\Delta \pi^{\frac{d}{2}} \Gamma(\frac{d}{2}-\Delta)}{\Gamma(\Delta)}
  \frac{\Gamma\bigl( \frac{1}{2}(\Delta+\ell) \bigr)^2 }{\Gamma\bigl( \frac{1}{2}(d-\Delta+\ell) \bigr)^2 }
\end{align}
\begin{figure}[h!]
    \centering
    \includegraphics[width=0.5\linewidth]{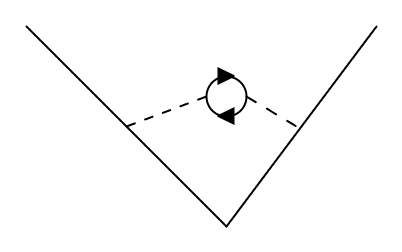}
    \caption{Scalar propagator with a fermion loop.} 
    \label{LS}
\end{figure}

The value of a scalar line with a loop on it as computed in flat space is (\ref{ss1l}).
If we naively convert this expression in flat space to the one on the cylinder, we have
\begin{equation}
    G_{flat\rightarrow cylinder}=\frac{g_1^2N}{64\pi^d}\frac{\Gamma(\frac{d}{2}-1)^2 \Gamma(d-3)}{\Gamma(d-2) (4-d)}\frac{1}{(2\cosh\tau-2\hat{n}_1\cdot\hat{n}_2)^{d-3}}
\end{equation}
Thus, if we compute the value of the diagram using this expression, we get
\begin{equation}
    GNY_{flat}=T\frac{h_1h_2g_1^2N}{64\pi^d}\frac{\Gamma(\frac{d}{2}-1)^2 \Gamma(d-3)}{\Gamma(d-2) (4-d)}\sum_{\ell,m}K_{d-3}(0,\ell)Y_{\ell m}(\hat{n}_1)Y_{\ell m}(\hat{n}_2)^{*}
\end{equation}
Now, let us turn our attention to the full calculation on the cylinder.
We have the expression given in (\ref{gnycyl}). 
Upon decomposing each piece of the form (\ref{gdeltdef}) into spherical harmonics, we get
\begin{equation}\label{gnyshdecomp}
    GNY=T \Big[-h_{1,0}h_{2,0}g_{1,0}^{2}\left(\frac{\Gamma(\frac{d}{2}-1)}{4\pi^{\frac{d}{2}}}\right)^4 (d-2)^2  N\Big]\sum_{\ell,m}K_{\frac{d}{2}-1}(0,\ell)^2K_{d-1}(0,\ell)Y_{\ell m}(\hat{n}_1)Y_{\ell m}(\hat{n}_2)^*
\end{equation}
We can now compare the expressions for $GNY$ and $GNY_{flat}$ as both are expansions in spherical harmonics. All we need to check is the relation between each coefficient.
Upon expanding around $d=4$, we get
\begin{equation}
    \frac{g_1^2N}{64\pi^d}\frac{\Gamma(\frac{d}{2}-1)^2 \Gamma(d-3)}{\Gamma(d-2) (4-d)}K_{d-3}(0,\ell) =\Bigg(\Big[-g_{1}^{2}\left(\frac{\Gamma(\frac{d}{2}-1)}{4\pi^{\frac{d}{2}}}\right)^4 (d-2)^2  N\Big]K_{\frac{d}{2}-1}(0,\ell)^2K_{d-1}(0,\ell)\Bigg)(1+\mathcal{O}(\epsilon^2))
\end{equation}
where $\epsilon=4-d$.
Thus, we have
\begin{equation}
    GNY_{flat}=GNY(1+\mathcal{O}(\epsilon^2))
\end{equation}
Since our calculation is to $\mathcal{O}(\epsilon)$, we can just use
\begin{equation}
    \begin{split}
       \frac{1}{64\pi^d}\frac{\Gamma(\frac{d}{2}-1)^2 \Gamma(d-3)}{\Gamma(d-2) (4-d)} \int dv \frac{e^{-(d-3)|v|}}{(1+e^{-2|v|}-2\hat{n}_1\hat{n}_2 e^{-|v|})^{d-3}}&=-\left(\frac{\Gamma(\frac{d}{2}-1)}{4\pi^{\frac{d}{2}}}\right)^4(d-2)^2\\&\int d\tau_1\int d\tau_2\int d\tau \int d^{d-1}\hat{n}_L\int d^{d-1}\hat{n}_R \\& \frac{\exp(-\frac{d-2}{2}|\tau_1|)}{(1-2\hat{n}_1\cdot\hat{n}_L e^{-|\tau_1|}+e^{-2|\tau_1|})^{\frac{d-2}{2}}}\\&\frac{\exp(-\frac{d-2}{2}|\tau_2|)}{(1-2\hat{n}_R\cdot\hat{n}_2 e^{-|\tau_2|}+e^{-2|\tau_2|})^{\frac{d-2}{2}}}\\&\frac{\exp(-(d-1)|\tau|)}{(1-2\hat{n}_L\cdot\hat{n}_R e^{-|\tau|}+e^{-2|\tau|})^{d-1}} \\&+\mathcal{O}(\epsilon^2)
    \end{split}
\end{equation}
Thus, we can simply use the flat space expression naively converted to the cylinder frame for our calculation. However, one would need to consider the general expression (\ref{gnyshdecomp}) if they wish to calculate the cusp anomalous dimension or associated quantities to higher order.

\subsection{Expansion of functions around $\theta=0,\pi$}\label{app:appA2}
In order to compute the Casimir energy and $C_D$, we need to know the expansions of the concerned functions around $\theta=0$ and $\theta=\pi$ respectively. We use the results already obtained in \cite{cuomo2024impuritiescuspgeneraltheory}.
\subsubsection*{Expansion around $\theta=0$}
 At small $\theta$, we have
\begin{equation}\label{smthpimth}
    \frac{\pi-\theta}{\sin\theta}= \frac{\pi}{\theta}-1+\dots
\end{equation}
\begin{equation}\label{smthi11}
    I_{11}(\cos\theta) =\frac{4\pi^5\log\theta}{\theta}+\frac{8\pi^5(1-\log2)}{\theta}+12\pi^4\log\theta+(4\pi^4\log64+24\pi^2\zeta(3)-32\pi^4)+\dots
\end{equation}
\begin{equation}\label{smthi12}
    I_{12}(\cos\theta)=\frac{\pi^7}{\theta}+16\pi^4\log\theta+(16\pi^4\log2+24\pi^2\zeta(3)-32\pi^4)+\dots
\end{equation}
\begin{equation}\label{smthf0p}
    F'_0(\cos\theta)=\frac{2\pi\log\theta}{\theta}+\frac{2\pi\log2}{\theta}-2+\dots
\end{equation}
Here, note that the only difference from \cite{cuomo2024impuritiescuspgeneraltheory} is in (\ref{smthf0p}). The function we use is, by definition, twice the function used in their calculation.
\subsubsection*{Expansion around $\theta=\pi$ }
We have
\begin{equation}\label{thpipimth}
    \left(\frac{\pi-\theta}{\sin\theta}-1\right)=\frac{1}{6}(\pi-\theta)^2 + \mathcal{O}((\pi-\theta)^4)
\end{equation}
\begin{equation}\label{thpif0p}
    F'(\cos\theta)-2=\frac{1}{9}(\pi-\theta)^2+\mathcal{O}((\pi-\theta)^4)
\end{equation}
\begin{equation}\label{thpii11}
    I_{11}(\cos\theta)-8\pi^4+4\pi^4\log8+24\pi^2\zeta(3)=\frac{4\pi^4-2\pi^2-2\pi^4\log8-12\pi^2\zeta(3)}{3}(\pi-\theta)^2+\mathcal{O}((\pi-\theta)^4)
\end{equation}
\begin{equation}\label{thpii12}
    I_{12}(\cos\theta)-24\pi^2\zeta(3) = (\sigma_0-24\pi^2\zeta(3))+\sigma_2(\pi-\theta)^2+\mathcal{O}((\pi-\theta)^4)
\end{equation}
Here,
\begin{equation}\label{sig0}
    \begin{split}
        \sigma_0&=16 \pi ^4 \log (2)-\frac{\pi ^6}{9}+
\Sigma_0
    \end{split}
\end{equation}
\begin{equation}\label{sig2}
    \sigma_2=\frac{1}{18} \pi ^2 \left[3 \pi ^4+48+4 \pi ^2 (16 \log 2-9)\right]+\Sigma_2
\end{equation}
\begin{equation}\label{Sig0}
    \Sigma_0=2 \pi ^2\sum_{\ell=1}^{\infty}(-1)^{\ell}\left\{\left[\pi ^2-2 \psi ^{(1)}\left(\frac{\ell+2}{2}\right)\right]^2-\pi ^4+\frac{8 \pi ^2}{\ell}\right\}
\end{equation}
\begin{equation}\label{Sig2}
    \Sigma_2=4 \pi ^2 \sum_{\ell=1}^{\infty}(-1)^{\ell}\frac{3 (\ell+2) \ell^2 \psi ^{(1)}\left(\frac{\ell+2}{2}\right) \left[\pi ^2-\psi ^{(1)}\left(\frac{\ell}{2}+1\right)\right]+12 \ell-2 \pi ^2 \left[3 \ell (\ell+1)-4\right]}{9 \ell}
\end{equation}
These results are already derived in \cite{cuomo2024impuritiescuspgeneraltheory}. However, we can circumvent the problem of non-commuting of the Taylor expansion of the the integrand of $I_{12}(\cos\theta)$ and the $\phi_x$ integral by subtracting an appropriate integral that leaves the quadratic term unchanged.

The zeroeth order term in the expansion of $I_{12}(\cos\theta)$ comes out to $24\pi^2\zeta(3)$. This matches the value of the integral at $\theta=\pi$. To compute the next term i.e. the coefficient of $(\pi-\theta)^2$, we subtract $I_{12}^{(s)}$ where
\begin{equation}
    I_{12}^{(s)}=\int d^{3}\hat{n}f_{4}^2(\hat{n}\cdot\hat{n}_1)
\end{equation}
We can directly calculate this integral and we obtain
\begin{equation}
    I_{12}^{(s)}=\frac{4\pi^4}{3}
\end{equation}
Thus, subtracting this term from the full integral should not change the coefficient of the quadratic term. We can now evaluate the difference of the integrals and we obtain
\begin{equation}
    I_{12}(\cos\theta)=\dots+\frac{8\pi^2}{9}(1+6\zeta(3))(\pi-\theta)^2+\dots
\end{equation}
The numerical value matches the value of $\sigma_2$ derived in \cite{cuomo2024impuritiescuspgeneraltheory}. Furthermore, this value allows the expression we obtain for $\mathcal{N}_D^2$ and $\mathcal{N}_t^2$ to match those obtained in \cite{barrat2025linedefectcorrelatorsfermionic,Gimenez_Grau_2022}

\section{2-Loop calculation of fixed points}\label{app:appB}
The beta functions of the bulk coupling constants can be extracted from the literature  \cite{K_rkk_inen_1994,Machacek:1983fi,MACHACEK198383,MACHACEK198570,VANDAMME1984105,ROSENSTEIN1993381}, and in our conventions they read
\begin{equation}
    \beta_{g_1}=-\frac{\epsilon g_1}{2} +\frac{N+8-2N_s}{2(4\pi)^2}g_1^3 + \frac{1}{(4\pi)^4}\Big(\frac{(N_s+2)}{36}g_1g_2^2-\frac{2(N_s+2)}{3}g_1^3g_2-\frac{(9N_S^2-40N_s+12N+40)}{4}g_1^5\Big)
\end{equation}
\begin{equation}
    \begin{split}
        \beta_{g_2}= -\epsilon g_2 &+ \frac{1}{(4\pi)^2}\Big(\frac{N_s+8}{3}g_2^2+2Ng_1^2g_2-12Ng_1^4\Big) \\ &+\frac{1}{(4\pi)^4}\Big(96Ng_1^6+(12-5N_s)Ng_1^4g_2-\frac{(N_s+8)N}{3}g_1^2g_2^2-\frac{14+3N_s}{3}g_2^3\Big)
    \end{split}
\end{equation}

The renormalization of the defect coupling to two-loop order receive contributions from the following diagrams 
\begin{figure}[h!]
    \centering
    \begin{subfigure}[b]{0.3\textwidth}
         \centering
         \includegraphics[width=\textwidth]{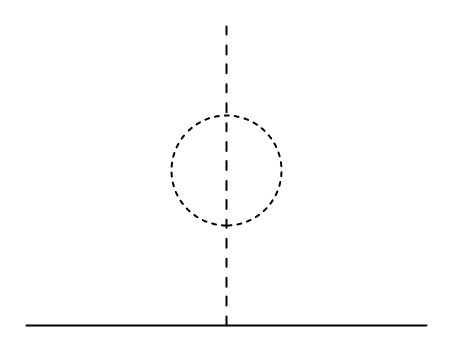}
         \caption{$g_2^2h$}
         \label{g22_1}
    \end{subfigure}
    \begin{subfigure}[b]{0.3\textwidth}
         \centering
         \includegraphics[width=\textwidth]{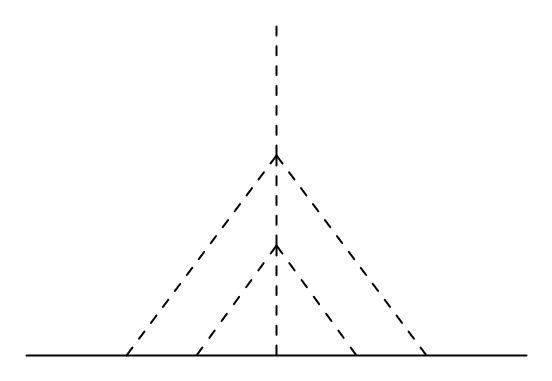}
         \caption{$g_2^2h^5$}
         \label{g22_2}
    \end{subfigure}
    \begin{subfigure}[b]{0.3\textwidth}
         \centering
         \includegraphics[width=\textwidth]{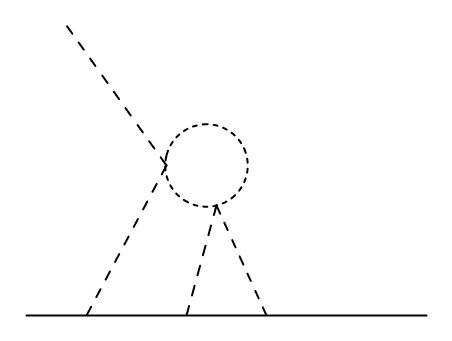}
         \caption{$g_2^2h^3$}
         \label{g22_3}
    \end{subfigure}
    \caption{$\sim g_2^2$}
    \label{g22}
\end{figure}

\begin{figure}[h!]
    \centering
    \begin{subfigure}[b]{0.3\textwidth}
         \centering
         \includegraphics[width=\textwidth]{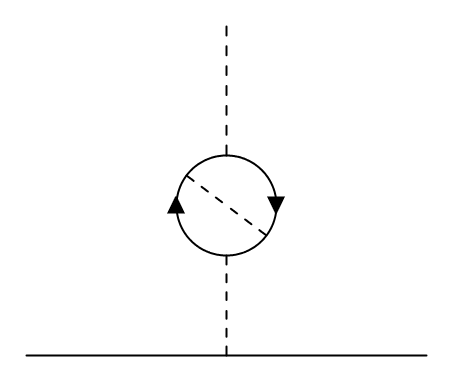}
         \caption{$g_1^4h$}
         \label{g14_1}
    \end{subfigure}
    \begin{subfigure}[b]{0.3\textwidth}
         \centering
         \includegraphics[width=\textwidth]{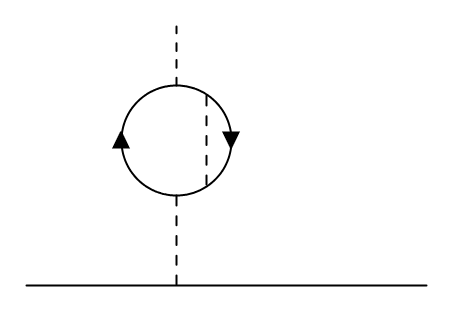}
         \caption{$g_1^4h$}
         \label{g14_4}
    \end{subfigure}
    \begin{subfigure}[b]{0.3\textwidth}
         \centering
         \includegraphics[width=\textwidth]{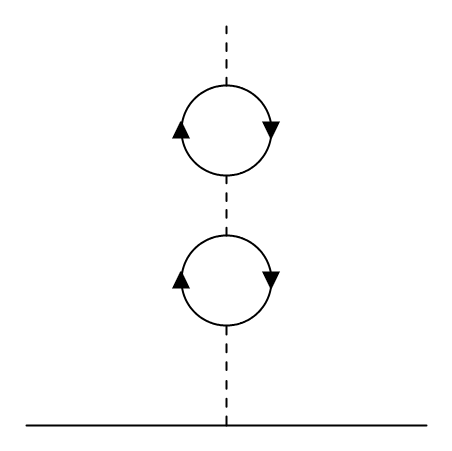}
         \caption{$g_1^4h$}
         \label{g22_2}
    \end{subfigure}
    \begin{subfigure}[b]{0.3\textwidth}
         \centering
         \includegraphics[width=\textwidth]{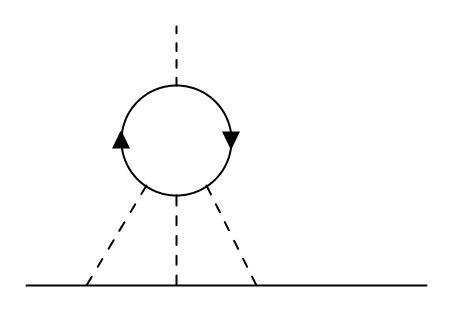}
         \caption{$g_1^4h^3$}
         \label{g14_3}
    \end{subfigure}
    \caption{$\sim g_1^4$}
    \label{g14}
\end{figure}

\begin{figure}[h!]
    \centering
    \begin{subfigure}[b]{0.45\textwidth}
         \centering
         \includegraphics[width=\textwidth]{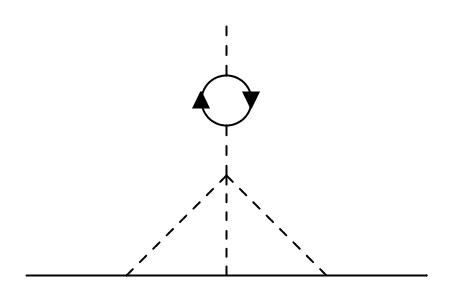}
         \caption{$g_2g_1^2h^3$}
         \label{g2g12_1}
    \end{subfigure}
    \begin{subfigure}[b]{0.45\textwidth}
         \centering
         \includegraphics[width=\textwidth]{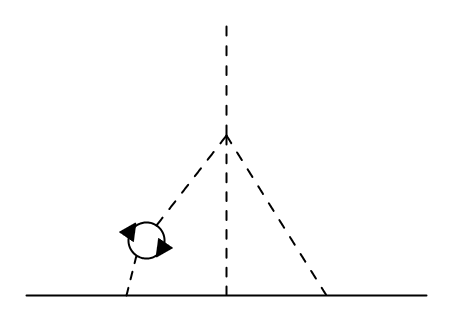}
         \caption{$g_2g_1^2h^3$}
         \label{g2g12_2}
    \end{subfigure}
    
    \caption{$\sim g_2g_1^2$}
    \label{g2g12}
\end{figure}

For the diagrams involving just the scalar field shown in figure \ref{g22}, the results coincide with those in \cite{Cuomo_2022}. The additional diagrams have been computed in \cite{Pannell_2023}. We find the beta function for $h$ to be 
\begin{align}\label{eq:betafcn-h}
\beta_h =& -\frac{\epsilon h}{2} + \frac{1}{(4\pi)^2} \left(\frac{g_1^2Nh}{2}+\frac{g_2 h^3}{6}\right) + \frac{1}{(4\pi)^4} \Bigg\{g_2^2 h \left(\frac{(2 + N_s)}{36} - \frac{h^2 (N_s+8)}{36} - \frac{h^4}{12}\right) - \frac{g_2 g_1^2 h^3 N}{4} \notag \\
 & + g^4_1 h \left(- \frac{(N_s+4)N}{4} + h^2 N \left(1 - \frac{\pi^2}{6}\right)\right)\Bigg\}
\end{align}

From the beta functions above, we can compute the bulk fixed point values as
\begin{equation}
    g_{1*}^2=g_{1*,0}^2\epsilon+g_{1*,1}^2\epsilon^2
\end{equation}
With
\begin{equation}
    g_{1*,0}^2=(4\pi)^2\frac{1}{N+8-2N_s}
\end{equation}
\begin{equation}
\begin{split}
    g_{1*,1}^2=\frac{4\pi^2}{(8+N_s)^2(8+N-2N_s)^3}\Big(&N^2 \big(2 + N_s\big) \big(-1 + \xi\big) \\
&\quad + 4 N \big(328 + 58 N_s + N_s^2 + 32 \xi + 20 N_s \xi + 2 N_s^2 \xi\big) \\
&\quad + 2 \Big(9 N_s^4 + N_s^3 \big(94 - 10 \xi\big) + 192 N_s \big(-9 + \xi\big) \\
&\quad\quad - 12 N_s^2 \big(5 + 3 \xi\big) + 64 \big(47 + 7 \xi\big)\Big)
\end{split}
\end{equation}
\begin{equation}
    g_{2*}=g_{2*,0}\epsilon + g_{2*,1}\epsilon^2
\end{equation}
With
\begin{equation}
    g_{2*,0}=\frac{24 \pi^2
\Big( 8 - N - 2 N_s + \sqrt{ N^2 + 4 (-4 + N_s)^2 + 4 N (28 + 5 N_s) } \Big)
}
{ \Big( 8 + N - 2 N_s \Big) \Big( 8 + N_s \Big) }
\end{equation}
\begin{equation}
    \begin{split}
        g_{2*,1}=-\frac{
24 \pi^2 
}{
\big(8 + N - 2 N_s\big)^4 \big(8 + N_s\big)^3 
\xi
}\Big[&
N^4 \big(20 + 6 N_s + N_s^2\big) \big(-1 + \xi\big) \\
&\quad - 48 \big(-4 + N_s\big)^4 \big(14 + 3 N_s\big) \big(1 + \xi\big) \\
&\quad - 6 N \Big(
N_s^5 \big(17 + 5 \xi\big) + 8 N_s^4 \big(46 + 5 \xi\big) + 
896 N_s \big(-19 + 7 \xi\big) \\
&\quad\quad - 16 N_s^3 \big(-78 + 19 \xi\big) + 
512 \big(33 + 23 \xi\big) - 64 N_s^2 \big(83 + 32 \xi\big)
\Big) \\
&\quad + N^2 \Big(
-48 N_s^2 \big(-62 + 5 \xi\big) + N_s^4 \big(11 + 25 \xi\big) - 
128 \big(229 + 29 \xi\big) \\
&\quad\quad - 64 N_s \big(-19 + 199 \xi\big) + 
N_s^3 \big(404 + 436 \xi\big)
\Big) \\
&\quad - N^3 \Big(
1424 - 464 \xi + N_s^3 \big(5 + 7 \xi\big) \\&+ 
2 N_s^2 \big(-7 + 67 \xi\big) + N_s \big(-20 + 548 \xi\big)
\Big)
\Big]
    \end{split}
\end{equation}
These match the corresponding expressions obtained for the GNY ($N_s=1$) and NJLY ($N_s=2$) models in \cite{fei2017yukawacftsemergentsupersymmetry}. In the large $N$ limit, one has
\begin{equation}
    g_{1*}^2=\frac{16\pi^2}{N}\epsilon + \mathcal{O}\Big(\frac{1}{N^2}\Big)
\end{equation}
\begin{equation}
    g_{2*}=\frac{192\pi^2}{N}+\mathcal{O}\Big(\frac{1}{N^2}\Big)
\end{equation}

From the beta function for $h$, we obtain the defect fixed point to be
\begin{equation}
    h_{*}^2=h_{*,0}^2+h_{*,1}^2\epsilon
\end{equation}
With
\begin{equation}
\begin{split}
    h_{*,0}^2=\frac{4(4-N_s)(N_s+8)}{\Big( 8 - N - 2 N_s + \sqrt{ N^2 + 4 (-4 + N_s)^2 + 4 N (28 + 5 N_s) } \Big)}
\end{split}
\end{equation}
\begin{equation}
\begin{split}
h_{*,1}^2=&\frac{1}{
3 \big(8 + N - 2 N_s\big)^2 \big(8 + N_s\big) 
\xi  \big(
N \big(-1 + \xi\big) - 2 \big(-4 + N_s\big) \big(1 + \xi\big)
\big)^2
}\times\\&\Big[
-3 N^3 \big(304 - 4 N_s - 22 N_s^2 + N_s^3\big) \big(-1 + \xi\big) \\
&\quad + 48 \big(-4 + N_s\big)^4 \big(170 + 45 N_s + 4 N_s^2\big) \big(1 + \xi\big) \\
&\quad - N^2 \Big(
N_s^4 \big(75 - 39 \xi + 8 \pi^2 \xi\big) + 
48 N_s^2 \big(246 + 15 \xi + 16 \pi^2 \xi\big) \\
&\quad\quad - 64 N_s \big(759 - 597 \xi + 32 \pi^2 \xi\big) + 
4 N_s^3 \big(777 - 231 \xi + 40 \pi^2 \xi\big) \\
&\quad\quad - 128 \big(1581 - 471 \xi + 128 \pi^2 \xi\big)
\Big) \\
&\quad + 2 N \Big(
N_s^5 \big(183 - 81 \xi + 8 \pi^2 \xi\big) + 
16 N_s^3 \big(-1353 + 480 \xi + 8 \pi^2 \xi\big) \\
&\quad\quad + 8 N_s^4 \big(-45 - 114 \xi + 16 \pi^2 \xi\big) - 
128 N_s \big(-387 + 1335 \xi + 64 \pi^2 \xi\big) \\
&\quad\quad - 64 N_s^2 \big(2415 - 576 \xi + 80 \pi^2 \xi\big) + 
512 \big(5019 - 159 \xi + 128 \pi^2 \xi\big)
\Big)
\Big]
\end{split}
\end{equation}
Here, we have defined
\begin{equation}
    \xi=\sqrt{\frac{N^2+4(N_s-4)^2+4N(28+5N_s)}{(8+N-2N_s)^2}}
\end{equation}

At large $N$, this becomes
\begin{equation}
    h_*^2=\frac{4-N_s}{2}+\epsilon\frac{(4-N_S)\pi^2+(24-9N_s)}{24} +\mathcal{O}\Big(\frac{1}{N}\Big)
\end{equation}

\appendix

\bibliography{refs}
\bibliographystyle{utphys}

\end{document}